\documentclass[pdflatex,sn-nature]{sn-jnl}

\usepackage{graphicx}%
\usepackage{multirow}%
\usepackage{amsmath,amssymb,amsfonts}%
\usepackage{amsthm}%
\usepackage{mathrsfs}%
\usepackage[title]{appendix}%
\usepackage{xcolor}%
\usepackage{textcomp}%
\usepackage{manyfoot}%
\usepackage{booktabs}%
\usepackage{algorithm}%
\usepackage{algorithmicx}%
\usepackage{algpseudocode}%
\usepackage{listings}%

\usepackage{subcaption}
\usepackage{tabularx}
\usepackage{natbib}
\usepackage{url} 
\usepackage[utf8]{inputenc}  

\usepackage{diagbox}
\usepackage[export]{adjustbox}  
\usepackage{tikz}  

\newcommand{\errModule}{\textsc{Err}}
\usepackage{xspace}
\newcommand{\AIRT}{AIRT\xspace}
\newcommand{\Nval}{62\xspace}
\newcommand{\Ntest}{60\xspace}

\theoremstyle{thmstyleone}%
\theoremstyle{thmstyletwo}%

\theoremstyle{thmstylethree}%

\def\CC{{C\nolinebreak[4]\hspace{-.05em}\raisebox{.4ex}{\tiny\bf ++}}}

\usepackage{relsize}

\begin{document}

\title[Article Title]{AI End-to-End Radiation Treatment Planning Under One Second}



\author[1]{\fnm{Simon} \sur{Arberet}}\email{simon.arberet@siemens-healthineers.com}
\author[1]{\fnm{Riqiang} \sur{Gao}}
\author[2]{\fnm{Martin} \sur{Kraus}}
\author[2]{\fnm{Florin} \sur{C. Ghesu}}
\author[3]{\fnm{Wilko} \sur{Verbakel}}
\author[2]{\fnm{Mamadou} \sur{Diallo}}
\author[3]{\fnm{Anthony} \sur{Magliari}}
\author[1]{\fnm{Venkatesan} \sur{Karuppusamy}}
\author[3]{\fnm{Sushil} \sur{Beriwal}}
\author{\fnm{REQUITE} \sur{Consortium}}
\author[1]{\fnm{Ali} \sur{Kamen}}
\author[1]{\fnm{Dorin} \sur{Comaniciu}}

\affil[1]{\orgdiv{Digital Technology and Innovation}, \orgname{Siemens Healthineers}, \orgaddress{\city{Princeton}, \state{NJ}, \country{USA}}}
\affil[2]{\orgdiv{Digital Technology and Innovation}, \orgname{Siemens Healthineers}, \orgaddress{\city{Erlangen}, \country{Germany}}}
\affil[3]{\orgdiv{Varian Medical Affairs}, \orgname{Siemens Healthineers}, \orgaddress{\city{Palo Alto}, \state{CA}, \country{USA}}}

\abstract{Artificial intelligence-based radiation therapy (RT) planning has the potential to reduce planning time and inter-planner variability, improving efficiency and consistency in clinical workflows.
Most existing automated approaches rely on multiple dose evaluations and corrections, resulting in plan generation times of several minutes.
We introduce \AIRT{} (Artificial Intelligence-based Radiotherapy), an end-to-end deep-learning framework that directly infers deliverable treatment plans from CT images and structure contours. \AIRT{} generates single-arc VMAT prostate plans, from imaging and anatomical inputs to leaf sequencing, in under one second on a single Nvidia A100 GPU.
The framework includes a differentiable dose feedback, an adversarial fluence map shaping, and a plan generation augmentation to improve plan quality and robustness.
The model was trained on more than 10,000 intact prostate cases. Non-inferiority to RapidPlan Eclipse was demonstrated across target coverage and OAR sparing metrics. Target homogeneity (HI = 0.10 $\pm$ 0.01) and OAR sparing were similar to reference plans when evaluated using AcurosXB. These results represent a significant step toward ultra-fast standardized RT planning and a streamlined clinical workflow.}




\maketitle


\section{Introduction}\label{sec1}

The main goals of automatic radiation therapy are to reduce planning time, improve inter-planner dose consistency, and maintain or enhance plan quality \cite{thompson2018artificial, hussein2018automation}.  
However, conventional inverse planning techniques rely on lengthy iterative optimization algorithms \cite{bortfeld2006imrt, shepard1999optimizing} and require substantial planner expertise \cite{nelms2012variation, hussein2018automation}.
These dependences prevent the availability of fast, consistent and high-quality treatment planning everywhere in the world, without the need for an experienced planner.

Most recent AI-based approaches for automatic radiation therapy planning are based on reinforcement learning (RL) \cite{dongrong2025breast, hrinivich2024clinical, mekki2025dual, gaomulti, shaffer2026AAPL, yang2025reinforcement} or model predictive control (MPC) \cite{yang2025foresight}.
These methods often involve multiple time-consuming iterative steps and/or interactions with the treatment planning system (TPS), resulting in planning times on the order of minutes. 
Some recent RL methods can achieve sub-minute inference time \cite{shaffer2026AAPL}, but then rely on additional preprocessing and coarse dose resolutions to remain tractable, which can limit their plan quality. In addition, RL-based methods typically decompose their planning into a sequence of local control-point steps, favoring computational tractability. This may come at the expense of global plan optimality.


In this work, we present a novel deep-learning (DL) method that can generate single-arc VMAT plans for prostate (without lymph nodes), including leaf sequencing, in under one second. Built around recent studies by the authors on AI for automatic radiation therapy planning \cite{arberet2025beam, gao2023flexible, kraus2025singleshot, gao2025generative, gao2025automating}, the method produces clinically deliverable plans directly from patient anatomy and achieves, for the first time, fully automated single-arc VMAT planning with a dosimetric quality similar to clinically used TPS optimization (RapidPlan Eclipse). The method learns from a large dataset of TPS plans while leveraging a differentiable dose-feedback mechanism that refines the fluence maps to improve clinically relevant objectives, particularly target dose homogeneity, all within an end-to-end pipeline with sub-second inference time.
The dose-feedback mechanism includes a DL dose computation internally, distinct from the final dose evaluation.
This mechanism improves plan quality relative to a simple feed-forward approaches \cite{mgboh2025fluenceformer, arberet2025beam} and improves planning speed compared to methods requiring multiple dose calculations \cite{dongrong2025breast, hrinivich2024clinical, mekki2025dual, gaomulti, shaffer2026AAPL, yang2025reinforcement}. 

This dose feedback mechanism also enables the implementation of a MCO-like approach where the user can manipulate a slider to vary the target homogeneity vs OAR sparing trade-off. However, this feature was only partially explored and is not the main focus of the present work.
By generating high-quality and consistent plans automatically, the present approach could improve access to advanced radiation therapy techniques in regions of the world where there is a scarcity of qualified dosimetrists and physicists.

\section{Related Work}\label{sec:related_work}

We can identify four directions of work in the field of AI-based automatic radiation treatment planning:
1) Reinforcement learning (RL) approaches that predict fluence maps of machine parameters \cite{gaomulti, dongrong2025breast,hrinivich2024clinical,mekki2025dual, shaffer2026AAPL};
2) RL methods that adjust the dose-volume objectives with the treatment planning system (TPS) in the loop \cite{yang2025reinforcement};
3) model-predictive control (MPC) that predict the plan quality and optimize objectives \cite{yang2025foresight};
4) direct approaches which predict fluence maps with a feed-forward network \cite{mgboh2025fluenceformer}.

Recent studies have demonstrated the feasibility of deep RL approaches for IMRT fluence map painting and machine parameter optimization, including single and dual-arc VMAT \cite{dongrong2025breast,hrinivich2024clinical,mekki2025dual}.
A recent study by the authors \cite{gaomulti} also uses deep RL for leaf-sequencing VMAT fluence maps, but without providing a robust TPS-free planning pipeline from CT imaging to final plan.
A very recent TPS-free RL method \cite{shaffer2026AAPL} reports a multi-second inference time. However it depends on additional pre-processing on the order of more than 20 seconds. 
Other works adjust objective constraints while invoking TPS after each step \cite{yang2025reinforcement}. 
MPC was also used to predict future dose responses and optimize discrete objective changes \cite{yang2025foresight}.
In this line of work, the treatment planning system (TPS) is used for iterative refinement and/or episodic rewards. When TPS warm-start and refinements are used, planning time ranges from 80 to 100 seconds. 
Beyond RL/MPC, other approaches explored learning-based dose and fluence prediction in order to accelerate planning. For example, transformer models have been used to predict fluence maps for multi-beam IMRT plans, but it does not close the loop with a differentiable dose feedback \cite{mgboh2025fluenceformer}. 
In another line of research, physics-informed differentiable dose engines enabled gradient-based optimization, but they still require an optimization loop to yield a complete clinical artifact \cite{simko2025physics} and require even longer optimization time than classical TPS optimization.

This approach differs by multiple aspects:
It is a full end-to-end pipeline without any TPS interaction. It contains a sequence of feed-forward modules for ultra-fast execution: auto-contouring, dose proposal, Bev2Fluence prediction, differentiable dose computation, single-pass dose-error correction, and finally leaf sequencing. 
One main contribution is a differentiable dose feedback mechanism which can optimize dose metrics accurately in a single feed-forward pass as opposed to multiple TPS refinements.
We also address sequencability using an adversarial loss on the fluence maps and an ad-hoc VMAT leaf-sequencer. 
The pipeline is fast and can generate single arc VMAT plans in \(<\)1\,s (including leaf-sequencing) which is faster, by multiple degrees of magnitude, to RL+TPS pipelines that require $\approx80$--$100$\,s including TPS refinement \cite{hrinivich2024clinical,mekki2025dual}.
By coupling a full AI feed-forward end-to-end pipeline with a differentiable dose-feedback mechanism before sequencing, this method addresses missing capabilities essential for ultra-fast, deployable VMAT planning.

\begin{figure}[H]
  \centering
  \includegraphics[width=0.99\textwidth]{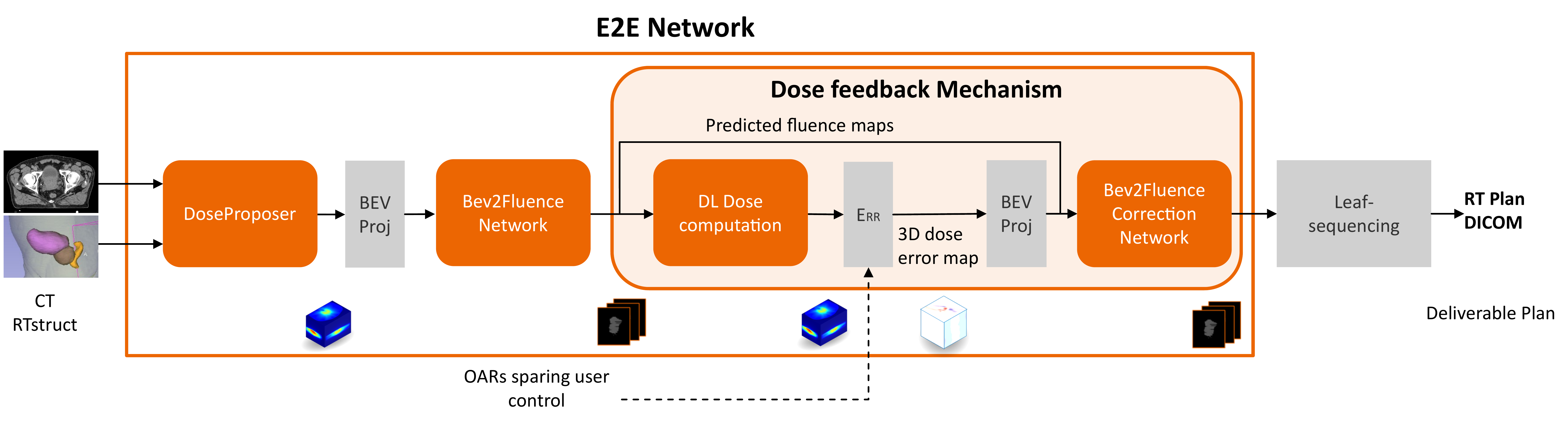}
  \caption{\AIRT end-to-end pipeline for AI VMAT plan generation.}
  \label{fig:end2end_pipeline}
\end{figure}

\section{Results}

\subsection{Overview}
The end-to-end AI-based VMAT planning pipeline (\AIRT) was evaluated on a dataset of intact prostate cases, against RapidPlan plans \cite{varian2024pelvisProstateSBRT, sackett2026sharing} optimized in Eclipse using the Photon Optimizer (PO) algorithm with AcurosXB (see Methods section \ref{sec:methods} for details). 
Performance was assessed in terms of dosimetric quality, statistical non-inferiority to clinical plans, computational efficiency, adaptability to user-defined organ sparing, and robustness across different patient anatomies.
For the dosimetric evaluations, multiple dose computation engines were employed: 1) a physics-informed DL engine~\cite{kraus2025singleshot} which is also integrated in the dose feedback mechanism (see Methods section), 2) an in-house LTBE solver (LTBS in short) derived from the AcurosXB codebase as described in~\cite{kraus2025singleshot},
3) the Eclipse AAA dose engine, 4) the Eclipse AcurosXB dose engine.  
Unlike Eclipse AcurosXB and AAA, the DL dose and LTBS dose engines are using a simplified source model. As a result, metrics may differ when evaluated with clinical dose engines (AAA and AcurosXB).
For conciseness, only results obtained with the DL dose engine and Eclipse AcurosXB are reported in the main manuscript, while results from all four dose engines are provided in the supplement material. 
The pipeline delivers VMAT plans in under one second per case on a single GPU, demonstrating that it is both clinically feasible and computationally efficient under steady-state inference conditions (i.e. after one-time model and GPU initialization).
A compute time breakdown analysis of our pipeline is available in Supplementary Table~S4. 
Figure~\ref{fig:end2end_pipeline} illustrates the full pipeline.

\subsection{Dosimetric Performance}

All dose metrics were evaluated using DL dose engine and AcurosXB (see above). 
It is important to note that AcurosXB, while widely considered a clinical refence standard for dose calculation, is itself a computational model and not the absolute ground truth.
Each planning method, when evaluated using its ``native" dose engine (\AIRT with DL dose, Eclipse with AcurosXB), will naturally make it look best. 

Plans generated by the \AIRT method were comparable in quality to those of the Eclipse plans.  
When evaluated with DL dose engine, \AIRT plans had lower target homogeneity index (HI) values (mean PTV HI = 0.11) compared to Eclipse plans (mean PTV HI = 0.16).
However this difference likely reflects the difference between the DL dose and AcurosXB rather than improvement in plan quality.

To unsure fair comparison, we also evaluated plans generated by both methods using the AcurosXB dose engine.
In this case, HI values were similar between the two methods even though the \AIRT method used DL dose in both its training and inference.
This demonstrates that \AIRT closely matches Eclipse optimized plan quality and suggests that further improvements may be achieved if the DL dose model was more closely aligned with AcurosXB.

Table~\ref{tab:comparison} shows dose-volume histogram (DVH) metrics of the \AIRT method vs Eclipse plans using AcurosXB. DVHs using the other dose engines are provided in the supplementary material.
Organs-at-risk (OARs) received comparable dose metrics with both planning methods. 
When evaluating with the DL dose engine, rectum doses were slightly lower for \AIRT{} plans while bladder doses were slightly higher. Similar trends were observed for the Rectum D50 when evaluated with AcurosXB.

\begin{table}[ht]
\centering
\small
\caption{Comparison of Eclipse and \AIRT{} planning for intact prostate VMAT cases.
All Eclipse plans were optimized using AcurosXB and \AIRT{} plans used the DL dose. For fair comparison, dose for both plan types, was calculated using the DL dose and AcurosXB engines.
The Homogeneity Index (HI) is defined as $(D_{2\%} - D_{98\%}) / D_{50\%}$ following the ICRU convention (lower is better). 
Data are presented as mean (SD). A * indicates $p < 0.05$ (Wilcoxon signed-rank test).}
\begin{tabular}{llcccc}
\hline
\multicolumn{2}{c}{\textbf{Dose Evaluation Engine $\rightarrow$}} & \multicolumn{2}{c}{\textbf{DL dose}} & \multicolumn{2}{c}{\textbf{Eclipse Acuros}} \\
\cmidrule(lr){3-4} \cmidrule(lr){5-6}
\textbf{Structure} & \textbf{Metric} & \textbf{Eclipse} & \textbf{AI} & \textbf{Eclipse} & \textbf{AI} \\
\hline
\multirow{2}{*}{\textbf{PTV}} & HI & 0.16\,(\scalebox{0.6}{0.03})* & 0.11\,(\scalebox{0.6}{0.02})* & 0.10\,(\scalebox{0.6}{0.01})* & 0.10\,(\scalebox{0.6}{0.01})* \\
 & D98 (Gy) & 38.4\,(\scalebox{0.6}{0.8})* & 38.8\,(\scalebox{0.6}{0.4})* & 39.3\,(\scalebox{0.6}{0.2}) & 39.3\,(\scalebox{0.6}{0.2}) \\
\hline
\multirow{3}{*}{\textbf{Bladder}} & Dmean (Gy) & 6.7\,(\scalebox{0.6}{3.3})* & 6.9\,(\scalebox{0.6}{3.5})* & 7.3\,(\scalebox{0.6}{3.5})* & 7.7\,(\scalebox{0.6}{3.8})* \\
 & D50 (Gy) & 2.9\,(\scalebox{0.6}{2.5}) & 3.0\,(\scalebox{0.6}{2.7}) & 3.3\,(\scalebox{0.6}{2.7})* & 3.6\,(\scalebox{0.6}{3.0})* \\
 & D2 (Gy) & 34.8\,(\scalebox{0.6}{6.8}) & 35.1\,(\scalebox{0.6}{6.6}) & 35.8\,(\scalebox{0.6}{6.0})* & 37.0\,(\scalebox{0.6}{5.9})* \\
\hline
\multirow{3}{*}{\textbf{Rectum}} & Dmean (Gy) & 5.5\,(\scalebox{0.6}{1.4})* & 5.3\,(\scalebox{0.6}{1.4})* & 5.4\,(\scalebox{0.6}{1.4}) & 5.5\,(\scalebox{0.6}{1.4}) \\
 & D50 (Gy) & 2.7\,(\scalebox{0.6}{1.2})* & 2.5\,(\scalebox{0.6}{1.1})* & 2.7\,(\scalebox{0.6}{1.0})* & 2.6\,(\scalebox{0.6}{1.0})* \\
 & D2 (Gy) & 30.1\,(\scalebox{0.6}{5.5})* & 30.5\,(\scalebox{0.6}{5.8})* & 31.4\,(\scalebox{0.6}{5.6})* & 32.5\,(\scalebox{0.6}{6.0})* \\
\hline
\end{tabular}
\label{tab:comparison}
\end{table}

\subsection{Statistical Validation and Non-Inferiority}


To evaluate statistical equivalence between the \AIRT method and Eclipse planning, we performed a non-inferiority test on the \Nval validation cases of the dataset using Eclipse AcurosXB dose engine.
For this test, we used a margin of 0.01 for the homogeneity index (HI), and a margin of 1.5 Gy to all other dose metrics.
Among the tested DVH metrics, the \AIRT method met the non-inferiority test margins at $p<0.05$. 
Details results for each DVH metric are provided in Supplementary Table~S3.

%

\subsection{Qualitative Results: Representative Cases}



\begin{figure}[H]
\centering
\begin{subfigure}{0.49\textwidth}
    \includegraphics[width=0.95\textwidth, clip, trim=0cm 0cm 5.8cm 0.9cm]{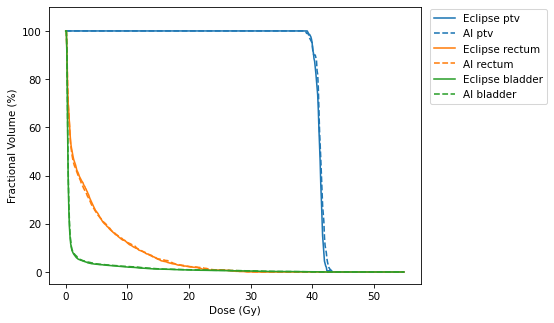} %
    \caption{Case 1}
    \label{fig:dvh1}
\end{subfigure}
\hfill
\begin{subfigure}{0.49\textwidth}
    \includegraphics[width=0.95\textwidth, clip, trim=0cm 0cm 5.8cm 0.9cm]{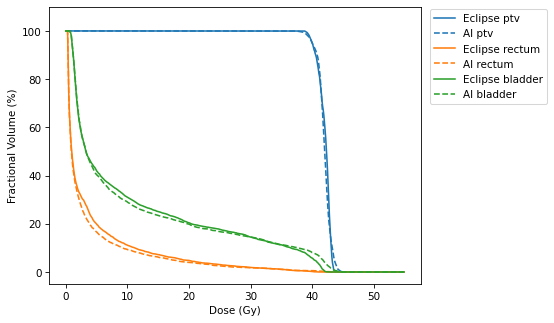}
    \caption{Case 2}
    \label{fig:dvh2}
\end{subfigure}

\begin{subfigure}{0.49\textwidth}
    \includegraphics[width=0.95\textwidth, clip, trim=0cm 0cm 5.8cm 0.9cm]{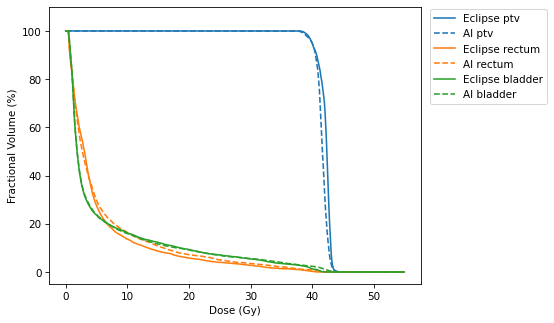}
    \caption{Case 3}
    \label{fig:dvh3}
\end{subfigure}
\hfill
\begin{subfigure}{0.49\textwidth}
    \includegraphics[width=0.95\textwidth, clip, trim=0cm 0cm 5.8cm 0.9cm]{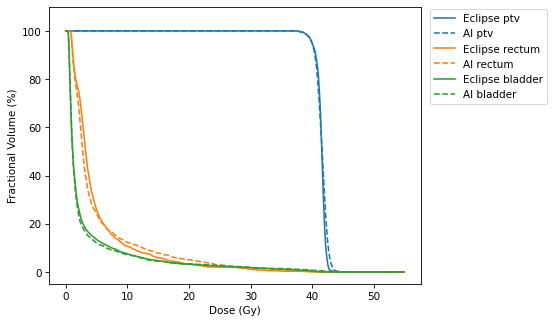}
    \caption{Case 4}
    \label{fig:dvh4}
\end{subfigure}

\begin{subfigure}{0.49\textwidth}
    \includegraphics[width=0.95\textwidth, clip, trim=0cm 0cm 5.8cm 0.9cm]{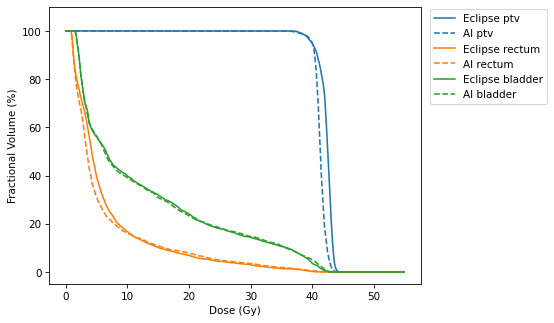}
    \caption{Case 5}
    \label{fig:dvh5}
\end{subfigure}
\hfill
\begin{subfigure}{0.49\textwidth}
    \includegraphics[width=0.95\textwidth, clip, trim=0cm 0cm 5.8cm 0.9cm]{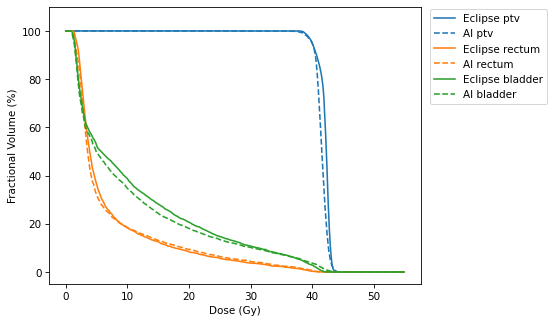}
    \caption{Case 6}
    \label{fig:dvh6}
\end{subfigure}

\vspace{-0.1em}

\centering
\begin{tikzpicture}
\node[
    draw,
    rectangle,
    rounded corners=2pt,
    line width=0.2pt,  
    inner sep=1pt
]{
\begin{tikzpicture}
\matrix[
    column sep=0.5em,  
    row sep=0.2em,  
    nodes={font=\footnotesize, anchor=west}
]{
    \draw[blue, line width=0.8pt] (0,0) -- (0.7,0); &
    \node{PTV (Eclipse)}; &
    \draw[blue, dashed, line width=0.8pt] (0,0) -- (0.7,0); &
    \node{PTV (AI)}; &
    \draw[orange, line width=0.8pt] (0,0) -- (0.7,0); &
    \node{Rectum (Eclipse)};  \\

    \draw[orange, dashed, line width=0.8pt] (0,0) -- (0.7,0); &
    \node{Rectum (AI)}; &
    \draw[green!60!black, line width=0.8pt] (0,0) -- (0.7,0); &
    \node{Bladder (Eclipse)}; &
    \draw[green!60!black, dashed, line width=0.8pt] (0,0) -- (0.7,0); &
    \node{Bladder (AI)}; \\
};
\end{tikzpicture}
};
\end{tikzpicture}

\vspace{0.3em}
\caption{Dose–volume histograms (DVHs) for six cases corresponding to the 0th (minimum), 20th, 40th, 50th (median), 80th, and 100th (maximum) percentiles of PTV size in the validation dataset using Eclipse AcurosXB dose engine.}
\label{fig:dvh_6cases_Acuros}
\end{figure}
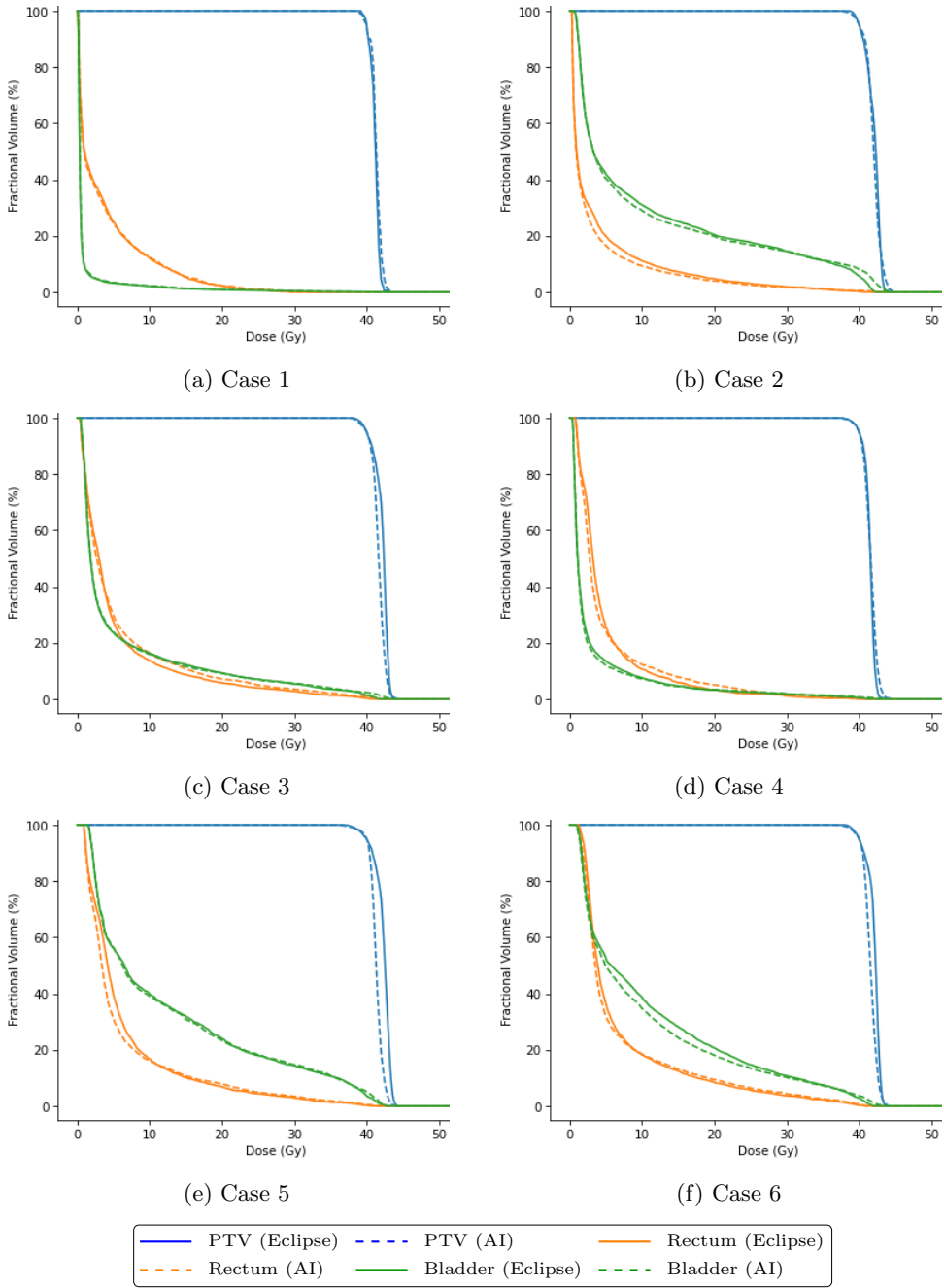

Figure~\ref{fig:dvh_6cases_Acuros} shows the DVH curves of the \AIRT method and Eclipse planning on six representative cases spanning the full range of PTV sizes of the validation dataset and evaluated using the Eclipse AcurosXB dose engine. 
To ensure an impartial and systematic selection process, avoiding cherry picking, we ranked all cases of the validation dataset by PTV volume sizes and selected the 0th (minimum), 20th, 40th, 50th (median), 80th, and 100th (maximum) percentiles. 
This approach provides both a representative coverage and an unbiased selection of cases.

Additionaly, for full transparency, we also provide the mean DVH curves across patients, as well as the results for every individual cases, for the \AIRT method and the Eclipse optimized plans in the Supplementary Material.
Results using other dose engine models (DL dose, LTBS, Eclipse AAA) are also provided in the supplementary material.
Visualization of the corresponding 3D dose distributions with the various dose engines are provided in supplementary Figures S15 to S22.
These results demonstrate that the \AIRT method maintained PTV coverage and OAR sparing over the full range of PTV sizes, showcasing the robustness of the \AIRT method to various anatomies.

\subsection{Fluence Map Quality and Leaf Sequencability}
\label{sec:fm_Q_leaf-seq}


In order to maintain the leaf sequenceability of the fluence maps, an adversarial loss was used during training. Supplementary Figure~S1 show some examples of fluence maps of the \AIRT method, compared to the fluence maps generated by the pipeline without adversarial loss, as well as the target fluence maps from Eclipse optimization.
The fluence maps generated with adversarial loss look more homogeneous and closer in distribution than the reference fluence maps.

The results from the ablation study in the next section also show that without adversarial loss, the PTV homogeneity is significantly worse, indicating issues in leaf sequencability if adversarial loss is not used. Note that all the dose metrics and DVH results reported in this study are based on generated plans after leaf sequencing.

\subsection{Ablation Study}


To isolate the contribution of the main components of the \AIRT{} method, we conducted an ablation study. In particular, we isolated: data augmentation (Aug), dose-feedback correction  (DF), and adversarial loss (GAN).
Across the variants, data augmentation systematically improved the results.
Dose feedback improved PTV homogeneity, but without adversarial loss, DF alone can also push the fluence maps away from the manifold of leaf-sequenced fluence maps and thus complicate the leaf-sequencability (see section \ref{sec:fm_Q_leaf-seq}). For this reason, the best results are obtained when DF and GAN are used together.
Note that when DF is used, relative metrics can get worse because DF focuses on improving absolute metrics (PTV homogeneity in this study), which can be in contradiction with the relative metrics. This is, for example, the case when the \AIRT plan has a PTV homogeneity better than the target.
\begin{table}[ht]
\centering
\resizebox{\textwidth}{!}{%
\begin{tabular}{lcccccccccc}
\toprule
 & \multicolumn{4}{c}{Relative metrics} & \multicolumn{6}{c}{Absolute metrics} \\
\cmidrule(lr){2-5} \cmidrule(lr){6-11}
 & \multicolumn{2}{c}{Fluence-domain} & \multicolumn{2}{c}{Dose-domain} & \multicolumn{2}{c}{PTV} & \multicolumn{2}{c}{Bladder} & \multicolumn{2}{c}{Rectum} \\
\cmidrule(lr){2-3} \cmidrule(lr){4-5} \cmidrule(lr){6-7} \cmidrule(lr){8-9} \cmidrule(lr){10-11}
Method & PSNR$\uparrow$ & SSIM$\uparrow$ & MAE$\downarrow$ & PTV MAE$\downarrow$ & HI$\downarrow$ & CI$\sim$ & mean$\downarrow$ & D2$\downarrow$ & mean$\downarrow$ & D2$\downarrow$ \\
\midrule
\AIRT (-Aug, -DF, -GAN) & 29.41 & 0.944 & 0.191 & 1.711 & 0.162 & 0.791 & 6.94 & 35.48 & \textbf{4.69} & \textbf{28.58} \\
\AIRT (-DF, -GAN) & \textbf{29.68} & 0.949 & 0.174 & 1.569 & 0.154 & 0.801 & \textbf{6.71} & 35.19 & 4.77 & 28.82 \\
\AIRT (-GAN) & 29.67 & \textbf{0.951} & \textbf{0.166} & 1.293 & 0.148 & 0.801 & 6.93 & 35.98 & 4.98 & 29.43 \\
\AIRT (-Aug) & 28.70 & 0.942 & 0.174 & \textbf{1.167} & 0.144 & 0.802 & 6.76 & 34.93 & 5.19 & 29.39 \\
\AIRT (full) & 29.09 & 0.945 & 0.167 & 1.575 & \textbf{0.107} & \textbf{0.822} & 6.85 & 35.11 & 5.29 & 30.51 \\
Eclipse Planning & \texttt{N.A.} & \texttt{N.A.} & \texttt{N.A.} & \texttt{N.A.} & 0.158 & 0.764 & 6.73 & \textbf{34.81} & 5.46 & 30.14 \\
\bottomrule
\end{tabular}
} 
\caption{
Ablation results for \AIRT variants, evaluated on \Nval cases of the validation set. Relative metrics are computed in comparison with target plans. Except for the fluence metrics (PSNR and SSIM), all the other metrics are dose-domain metrics, which are computed after leaf sequencing and using the DL dose engine~\cite{kraus2025singleshot}. Absolute DVH metrics are reported for PTV, bladder and rectum. 
``-Aug'' indicates that data augmentation was removed, ``-DF'' that dose feedback was removed, and ``-GAN'' that adversarial loss was removed.  ``full'' indicates that all the components of the AIRT method are used. Best values per metric are bold.
Units: Bladder and Rectum dose metrics are in Gy; PSNR is in dB; SSIM, HI and CI are unitless.
}
\label{tab:ablation}
\end{table}




\subsection{Adaptability: OARs Sparing Adjustments}

An additional benefit of the \AIRT framework is its flexibility.
As decribed in the Methods section, the dose feedback mechanism allows the user to optionally apply a scalar penalties to excess dose in OARs during inference.
While this functionality is not the main focus of the present work and is not yet fully explored, we include it here to demonstrate the flexibility of the \AIRT framework.
With appropriate training to accept scalar control inputs, the system enables per-patient dose adaptation for OAR sparing using simple user controls at inference time. 
For each organ at risk, the user can input a scalar value to increase the dose penalty in that organ.
This enables the generation of plans with various PTV-OAR trade-offs, without the need to retrain or replan.
This approach differs from the author's previous work \cite{gao2025generative}, which focused on dose prediction rather than direct generation of deliverable plans. 

\begin{figure}[H]
    \centering
    \includegraphics[width=.99\linewidth]{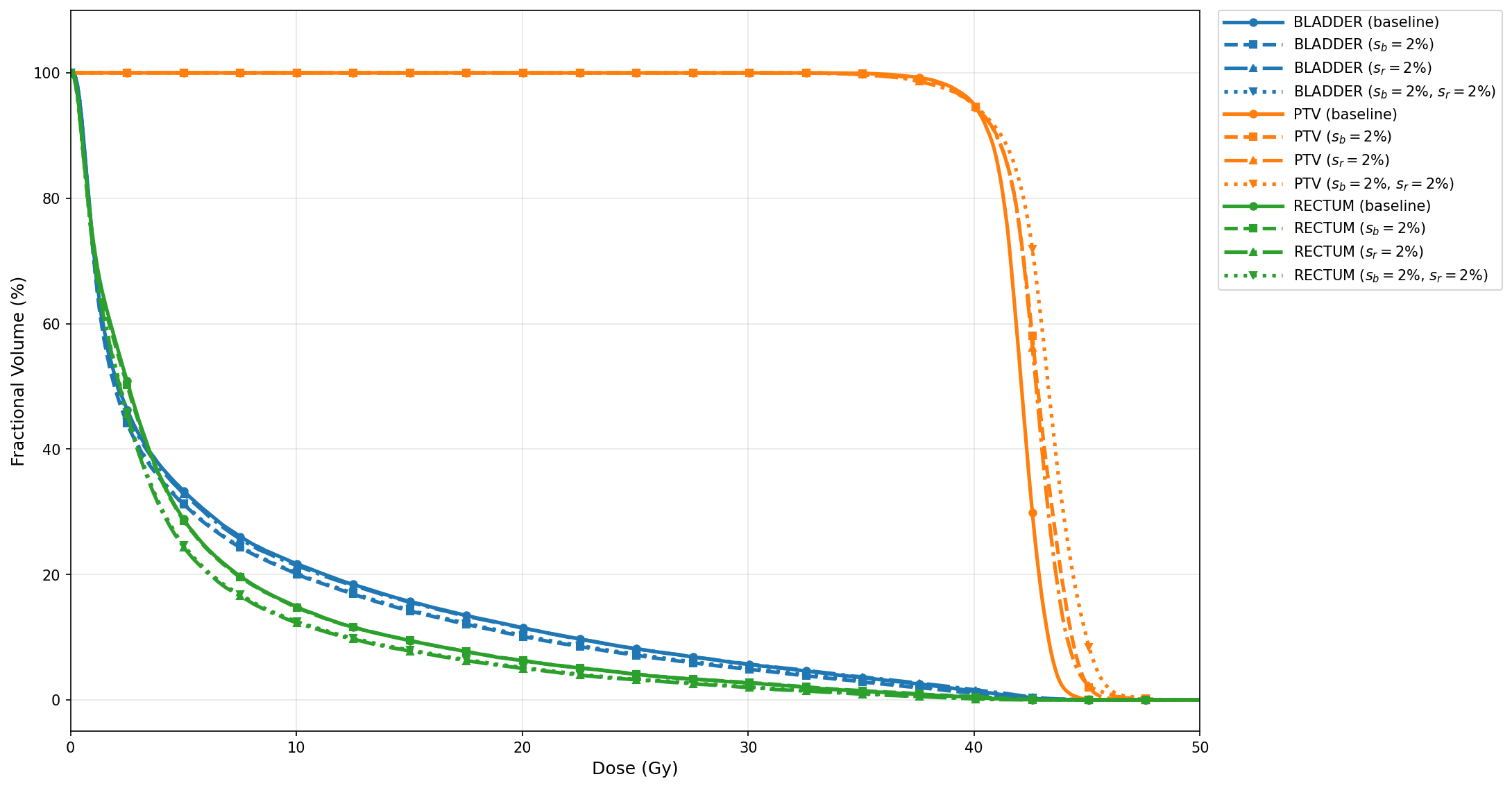}
    \caption{Averaged DVHs (over the \Nval cases of the validation dataset) of the \AIRT{} method for various OAR sparing controls. The ``(baseline)" planning, in the legend, means that no input OAR sparing control was used (equivalent to $s_r=s_b=0$). 
$s_b=2\%$ in the legend means that the dose feedback mechanism tried to decrease the dose in the bladder by $2\%$ voxel-wise (likewise for the rectum when $s_r=2\%$) compared to its input dose distribution.
}
    \label{fig:oar_adapt_dvh}
\end{figure}

To validate this adaptability, we generated all the plans of the validation dataset for different OAR scalar values for the bladder and rectum. 
Figure~\ref{fig:oar_adapt_dvh}, shows the averaged DVHs, over the validation dataset, evaluated using the Dl dose engine, of \AIRT for four different settings (no control, increased bladder sparing, increased rectum sparing, increased sparing on both bladder and rectum).
Additional DVH trade-offs are provided in the Supplementary Material. 
Results show that increasing the sparing factor of the bladder or rectum independently decreases the dose in the corresponding organ, with very limited effect on the other organ. 
However, as expected, increasing the OAR sparing on either of the organs, and even more on both, decreases the PTV homogeneity. This reflects the classical trade-off between PTV homogeneity and OAR sparing in radiation therapy.

\section{Discussion}

In this work, we show that an AI end-to-end pipeline can generate clinically deliverable single-arc VMAT plans from CT and contours in under one second. By combining a differentiable dose-feedback correction with an adversarial fluence loss, our pipeline produces fluence maps that can reliably be translated into MLC sequences. 
Trained on $>10{,}000$ intact-prostate plans generated in Eclipse using a RapidPlan model specifically developed for this indication~\cite{varian2024rapidplan, varian2024pelvisProstateSBRT, sackett2026sharing}, it achieves plan quality competitive with TPS baselines in under one second, avoiding the iterative TPS optimization loop (PO iterations). 
All results reported are obtained without warm-start refinements; however, the method may optionally be used as a warm-start for further manual refinement if desired.
This would extend the usability and scope of our approach by enabling the clinitian to customize plans for special cases or needs beyond those addressed by our AI pipeline.

Reducing planning times from minutes to under a second could transform clinical workflow. By generating standardized, clinically deliverable plans almost instantaneously, the pipeline can increase throughput in high-volume centers and deliver standardized plans learned from a large set of cases, with consistent planning quality.
This planning speed  enables an interactive mode in which clinicians can review multiple candidate plans in real time, discuss trade-offs, and converge on an acceptable deliverable plan within a single session. Because the pipeline produces DICOM RT Plans, it can easily be integrated into clinical systems.

In this work, we showed that we can achieve clinically deliverable fluence maps with a single feed-forward network, by using a differentiable dose error correction module, sufficient network capacity, and a large training set size. Unlike RL and MPC methods, which rely on multiple TPS calls and episodic rewards, our current pipeline maintains end-to-end differentiability during training, allowing dose objectives to be optimized via backpropagation through the different modules of the pipeline.
Adversarial loss enforces consistency with deliverable fluence maps and allows leaf-sequencing with a single shot without TPS refinements.
The ablation study demonstrated the importance of the dose-feedback combined with the adversarial training in improving target homogeneity.

Importantly, the dose-feedback mechanism itself is adaptable to multiple dose objectives: while we report results for a specific set of dose objectives chosen as proof-of-concept (e.g., reducing PTV inhomogeneity and possibly adding scalar penalty to OAR sparing), the same framework could be adapted to other differentiable dose objectives. 
This functionality presents an alternative to traditional Multi-Criteria Optimization (MCO), allowing clinitians to explore a range of trade-offs, adjusting objectives and viewing the results almost instantaneously, but with the difference that no replaning optimization is needed. 

Generalization is key to translating this method into clinical practice. While large-scale training and data augmentation improve generalization, as validated in our study,
it is essential to validate the performance of our method across various clinical institutions and anatomies. 
In this context, the American Oncology Institute (AOI) in India has independently tested our AI-generated plans, achieved target coverage comparable to manually crafted plans and successfully passed patient-specific quality assurance testing for deliverability.
These clinical results, currently unpublished, will be submitted as a conference abstract and a journal publication.

This study has some limitations. It is focused on single-arc VMAT intact prostate planning, and would require adaptations to address different body regions (Lung, Breast, Head \& Neck) or radiation therapy techniques such as multi-arc VMAT or simultaneous integrated boost (SIB).
The pipeline's architecture could be extended to multiple arcs or other more complex protocols, by adapting the network inputs/outputs, generating corresponding large scale datasets, and possibly using multiple dose-feedback loops to address the increased underdeterminacy of such problems. 
The current model was trained and validated for the Varian Millennium 120 (M120) MLC. 
Retraining may be necessary to adapt to other MLC models, especially if the leaf width differs, as the resulting fluence maps would have a different distribution. A flexible fundation model adressing multiple configurations could also be developped.

Our dose engine, while efficient and differentiable, is an approximation of clinical dose calculation. It relies on a simplified source model, with a spatially constant photon spectrum and without electron contamination effects. 
In contrast, AcurosXB is using a spatially varying spectrum and models secondary electrons.
The observed differences between the DL dose engine and AcurosXB, particularly with respect to the homogeneity index, suggests that a closer alignment of source and beam models between the two could further improve the overall quality of the AI-generated plans.
Interestingly, the AI generated plans showed less variation in homogeneity index across the different dose engines compared to Eclipse plans, suggesting that the AI method may be more robust to changes in dose modeling and potentially to the true clinical dose.
Future improvements will focus on more realistic source modeling to better match clinical solvers such as AcurosXB. 

Also, in order to facilitate rapid inference and manageable memory use, the deep-learning dose engine used in this study operates on a 4 mm grid resolution, which is slightly coarser than typical TPS dose engines. Progress in differentiable dose engine efficiency could enable higher dose resolution in our end-to-end pipeline, and potentially improve PTV-OAR sparing or in general, be able to address more targeted dose objectives (e.g. SIB).  

Future work includes generalizing our end-to-end pipeline to multiple arcs VMAT, addressing other body regions (Lung, Breast, Head \& Neck), as well as more complex and adaptable dose objectives. Overall, by eliminating TPS in-the-loop optimization while preserving deliverability, this pipeline points toward real-time radiotherapy planning that can complement current clinical workflows with faster iteration and consistent plan quality.

\section{Methods}
\label{sec:methods}
\subsection{Dataset and Augmentation}

The dataset consists of CT images, primarily from the REQUITE prostate dataset (1001 scans) and a few smaller in-house CT datasets, resulting in a total of 1,277 CT patient scans after curation.
The first step of the data creation process is the automatic segmentation of organs and helper structures, which are used by RapidPlan.
AI-Rad Companion Organs RT~\cite{AIRadCompanionOrgansRT} was used as the auto-contouring solution. Helper structures were generated via an in-house data curation pipeline implemented in \CC~\cite{gao2025automating}. 
The full list of contours and helper structures is depicted in the supplementary Table~S1. 
Then for each CT and its associated contour set (RT STRUCT), we created an RT plan in Eclipse with a publicly available RapidPlan model for prostate SBRT~\cite{varian2024pelvisProstateSBRT, sackett2026sharing}, 
 adapted for intact prostate (with PTV, CTV, and PTV  $\cap$ Rectum all set to 40Gy) and delivered as a single arc VMAT with a Varian Millennium 120 (M120) multileaf collimator (MLC).
Out of the 1,277 patients, 1,122 were randomly selected for training, \Nval for validation, and \Ntest for testing.

In order to increase network performance and generalization, we developed and applied a data-augmentation strategy to expand the training dataset from 1,122 training plans to 12,302.
The augmentation strategy includes the following transforms applied on the CT and contours: random image scaling, slight rotation, non-rigid deformation, and margin adjustments to PTV and rectum contours within clinically plausible ranges. All the augmented CT and their corresponding contours were then used to create new plans in Eclipse, following the same RapiPlan model.

\subsection{Pipeline Architecture}

The modular structure of the end-to-end pipeline is depicted in Figure~\ref{fig:end2end_pipeline}.
It takes the CT volume and contours (Body, PTV, OARs) as input and produces a leaf-sequences plan as output. 
This feed-forward design allows ultra-fast (\(<\)1\,s) generation of VMAT plan.
In the following, we detail each module of the pipeline.



\paragraph{Dose proposer}

The first module of the pipeline is the doseProposer, which takes the CT volume and contours (Body, PTV, OARs) as input and predicts a 3D dose distribution~\cite{gao2023flexible}. The architecture of this network is a 3D ResUNet, i.e. a U-Net encoder-decoder architecture with ResNet-style residual blocks and skip connections.

\paragraph{BEV projection (parameter-free)}

The 3D dose volume from the doseProposer is projected into the beam's eye view (BEV) of every control point (CP) of the VMAT plan, in order to place the dose in the same geometry as the fluence maps \cite{arberet2025beam}. 
This alignment simplifies the task of the Bev2Fluence network by creating a spatial correspondence between its input (dose BEV) and its output (fluence maps).

\paragraph{Bev2Fluence}

The Bev2Fluence network predicts the 180 fluence maps for the VMAT plan from its stack of 180 input beam's eye view dose projections~\cite{arberet2025beam}.
These fluence maps are predicted jointly due to the strong coupling between control points. 
This coupling is due to the VMAT delivery constraints (e.g. maximum leaf speed), but also because the resulting dose is the accumulation of every control point contribution.
The architecture of the Bev2Fluence network is a 3D convolutional network based on the MedNeXT backbone~\cite{roy2023mednext}.


\paragraph{Physics-informed dose computation}

In order to perform inference-time dose correction, a differentiable dose engine, which computes the dose from the input fluence maps, is used in the pipeline.
This dose engine is a deep-learning network using a physics-informed approach~\cite{kraus2025singleshot}.
Note that this dose engine was trained on an LTBE solver leveraging transport physics equations similar to AcurosXB, but employs a simplified beam model. 
Its architecture is based on a two-step design, where the fluence maps are first accumulated into 3d volume using spherical harmonics, together with the CT, and then an image-to-image network based on the MedNeXT backbone~\cite{roy2023mednext} is used to predict the final dose. 
Its inference time scales sub-linearly with the number of control points. For stability purposes, the network was trained~\cite{kraus2025singleshot} and is kept frozen.
We are using the 4 mm dose resolution version of this network in order to keep the inference time below one second.

\paragraph{Dose error module (parameter-free): \errModule}
%

This module generates a 3D dose error map by comparing the predicted dose and the desired dose distribution.
We focus primarily on the PTV homogeneity by using an asymmetrical metric that penalizes the hot spots more than the cold spots.
Optionally, this module can take \textit{scalar dose inputs}, to enable the user to penalize the excess dose in each OAR.
Each scalar input acts as a weight that scales how much dose need to be removed from a particular OAR in the baseline plan (i.e. the plan generated with a zero or no OAR scalar dose input).
By adjusting these values the clinician can interactively explore different trade-offs between target homogeneity and OAR sparing.
When this scalar is used, a 3D dose error component proportional to that weight is added to the 3D dose error map, which is then inputed to the Bev2Fluence correction module which role is to adjust the fluence maps to reduce that error.
The dose error framework is general and can, in principle, accommodate any differentiable dose metric (see details in Section~\ref{sec:dose-feedback-mechanism} and Supplementary Note~S1). 
This scalar dose input mechanism is one possible implementation.

The resulting dose error map is projected into the BEV before being provided to the Bev2Fluence correction network.

\paragraph{Bev2Fluence correction}

A second Bev2Fluence network refines fluence maps using as inputs: the initial fluence maps prediction and the BEV-projected dose error. 
This network, which is based on a similar architecture as the first Bev2Fluence but adapted to take additional inputs, corrects the input fluence maps using the BEV dose error as guidance. 
This module closes the dose feedback loop of the pipeline.

\paragraph{Leaf sequencing (rule-based)}
VMAT fluence exhibits a near two-level structure (open aperture vs. near-zero outside). 
Fluence maps are finally converted into leaf positions and monitor units with a rule-based leaf sequencer which models partial-pixel effects at moving leaf boundaries, and dosimetric leaf gap. 
The code is parallelized and implemented in Cython/C for speed.
Further implementation details are provided in Section~\ref{sec:leaf-sequencing}.


\paragraph{RT Plan export}
Finally, the leaf sequences and MU values can be exported into a DICOM RT Plan in order to be imported into a TPS such as Eclipse for review or direct delivery.
Alternatively, to streamline the workflow, the pipeline could be integrated into a TPS.

\subsection{Dose Feedback Mechanism}
\label{sec:dose-feedback-mechanism}

One of the main contributions of the \AIRT{} method is a fully differentiable dose feedback mechanism embedded in the architecture of our network. It allows the network to refine the fluence maps based on dose discrepancies, to improve target homogeneity and OAR sparing.
Indeed, the end-to-end pipeline begins with the DoseProposer network, which predicts a clinically plausible 3D dose distribution. From this dose, a first Bev2Fluence network infers 180 fluence maps. 
Coordinating these fluence maps so that their combined effect yields the desired dose distribution is a highly challenging task.
To tackle this challenge, the dose feedback mechanism operates as follows:
\begin{itemize}
	\item The dose distribution delivered by the initial predicted fluence maps is computed with the differentiable dose engine.
	\item A 3D dose error map is computed by the \errModule{} module based on the dose discrepancy with some dose metrics. This dose error map creation is very general and can, in principle, be instantiated by the gradient with respect to the dose of any differentiable dose planning objective. In this study, we implement it as a penalization of non-homogeneity within the PTV and, optionally, user-steered scalar penalties for excess dose in OARs. Futher details on this penalty impementation are provided in the Supplementary Material.
	\item This dose error map is projected into beam’s-eye view and combined with the initial fluence maps to feed the inputs of a second Bev2Fluence network, which is trained to produce a new set of fluence maps that reduce that dose error.  
\end{itemize}

In order to train the Bev2Fluence Correction module to reduce that dose error, the same error metric (produced by the Err module) is recomputed at training time, on the updated fluence maps, and a loss term penalizing the L1 loss of that final dose error map is added to the training optimizer.

This dose feedback correction could be applied multiple times, like an unrolled iterative gradient correction \cite{gregor2010learning}. 
However, in order to maintain computational efficiency, and as the results were already sufficient with a single pass, we used only one iteration of the dose feedback.
Without dose feedback, achieving target homogeneity and clinically acceptable dose distributions with a single prediction step would be extremely difficult, due to the complex coordination required among the large number of fluence maps.

\subsection{Adversarial Fluence Shaping}

When optimizing with a dose-based loss, the predicted fluence maps can deviate from the manifold of deliverable VMAT fluence maps, which have a characteristic two-level pattern that facilitates feasible leaf sequencing. In order to keep the fluence maps deliverable, we introduce an adversarial loss with a discriminator network that is trained to distinguish between fluence maps generated by the pipeline and target fluence maps generated from Eclipse-optimized plans. Additional technical details are provided in Supplementary Note S2.

\subsection{Training Strategy}

\begin{figure}[H]
  \centering
  \includegraphics[width=\linewidth]{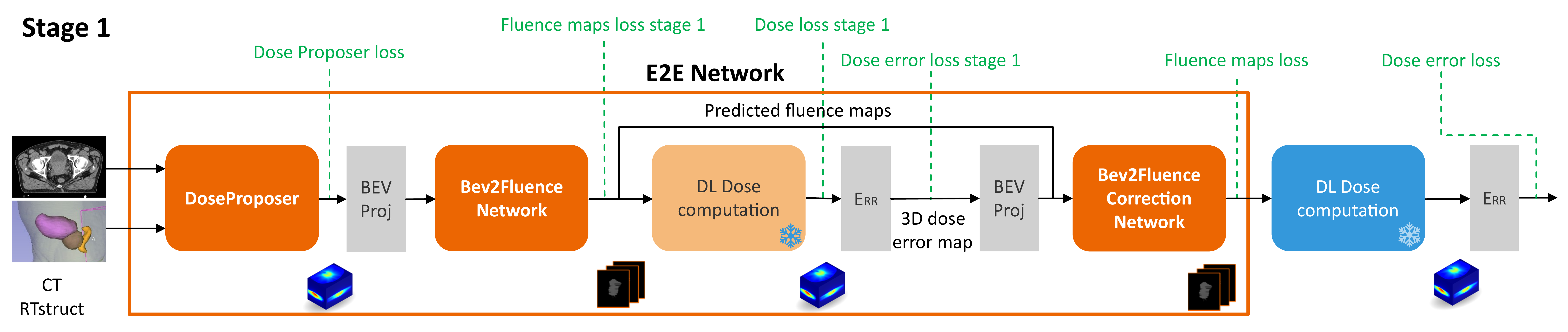}
  

\rule{\linewidth}{0.4pt}

  \includegraphics[width=\linewidth]{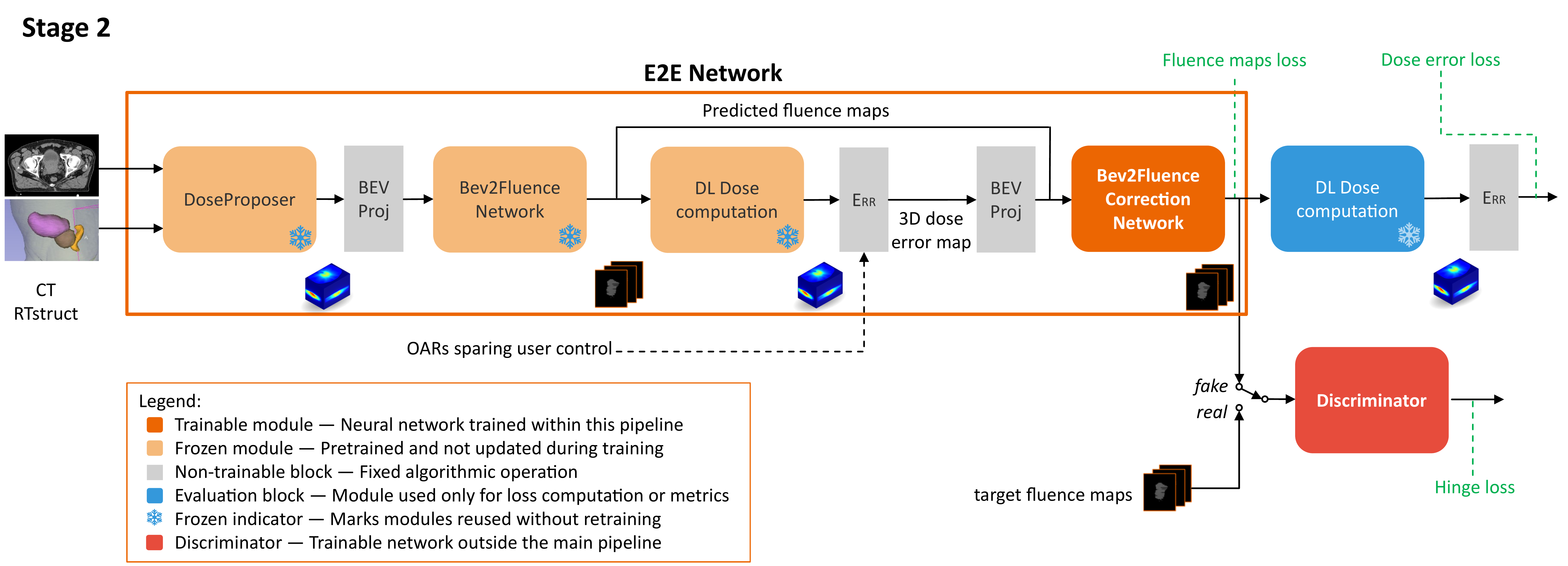}
  
  \caption{Two-stage training of the \AIRT end-to-end VMAT planning pipeline. Top: Stage 1 Full pipeline training without adversarial loss. Bottom: Stage 2: Fluence correction network training with adversarial loss.}
  \label{fig:training_pipeline}
\end{figure}

We train the AIRT network in two stages: (i) The first stage aims to train the pipeline to predict fluence maps under a stable loss function, and (ii) the second stage aims to enforce deliverable fluence maps and dose-feedback behavior under optional user-specified controls (Fig.~\ref{fig:training_pipeline}).

\textbf{Stage 1 (non-adversarial pretraining).}
We first train the full pipeline with reconstruction losses and without the adversarial loss.
The goal is to establish a stable training without the complications introduced by the deliverability constraints and the adversarial loss.
As depicted in Figure 4, the reconstruction losses include: 
\begin{itemize}
\item the ``Dose Proposer loss" and the ``Dose loss stage 1" which are L1 losses between predicted doses at different stages of the pipeline and the target dose, 
\item the ``Fluence maps loss stage 1" and ``Fluence maps loss", which are L1 losses between predicted fluence maps at different stages of the pipeline and the target fluence maps.
\item the ``Dose error loss stage 1" and the ``Dose error loss", which are L1 losses on the dose error maps at different stages of the pipeline. 
\end{itemize}

\textbf{Stage 2 (fluence correction with adversarial regularization and controls).}
In order to facilitate leaf-sequencability, an adversarial loss is introduced in the second stage of training. Its goal is to force the fluence maps to stay within the manifold of realistic VMAT fluence maps.
During that training stage, optional user-control parameters are randomly generated in a range consistent with their expected usage and passed to the two  \errModule{} modules (the one in the network pipeline and the one outside of the pipeline which is used for the ``Dose error loss" computation).
This ensures consistency between the dose error correction module and the dose loss function. 
For stability and efficiency reasons, the part of the network before the Bev2Fluence Correction module is frozen during that training stage.
See supplementary note~S2 for additional details.

\subsection{Leaf Sequencing}
\label{sec:leaf-sequencing}

In order to translate the fluence maps of the \AIRT{} pipeline into an RT Plan, we developed a rule-based leaf-sequencing algorithm adapted for VMAT plans.
The algorithm starts by converting the fluence maps of each control point (CP) into a discrete aperture, defined by the positions of the left and right leaves of each row of the multi-leaf collimator (MLC).
For each row of the MLC, the algorithm identifies the largest contiguous region of significant fluence in order to predict an initial aperture shape.
The Monitor Unit (MU), i.e.,  the radiation intensity of this control point,  is computed by averaging the fluence pixels in that initial aperture shape.
Some post-processing steps refine the positions of the leaves to: 1) account for the fact that leaf positions can be located in sub-pixel locations, leading to partially exposed pixels, and 2) account for the dosimetric leaf gap.

To efficiently process the 180 fluence maps of the VMAT plans, the leaf-sequencing algorithm is optimized for speed, runs in parallel for each control point, and is implemented in C.

\backmatter

\bmhead{Supplementary information}

Supplementary material is available for this paper.

\bmhead{Disclaimer}
The concepts and information presented in this paper/presentation are based on research results that are not commercially available. Future commercial availability cannot be guaranteed.

\bmhead{Acknowledgements}

We thank James Robar, Medical Physicist and Professor of Radiation Oncology at Nova Scotia Health Authority, for his valuable clinical insight and feedback.
We also thank Laura Balascuta, Liang Gao and Ioan-Marius Popdan for their help with data curation, ESAPI scripting and dose computation in Eclipse.


We thank all the contributors to the REQUITE project, including the patients, clinicians and nurses. The core REQUITE consortium consists of David Azria, Erik Briers, Jenny Chang-Claude, Alison M. Dunning, Rebecca M. Elliott, Corinne Faivre-Finn, Sara Guti\'errez-Enriquez, Kerstie Johnson, Zoe Lingard, Tiziana Rancati, Tim Rattay, Barry S. Rosenstein, Dirk De Ruysscher, Petra Seibold, Elena Sperk, R. Paul Symonds, Hilary Stobart, Christopher Talbot, Ana Vega, Liv Veldeman, Tim Ward, Adam Webb and Catharine M.L. West. \\

\newif\ifnature
\naturefalse   
\ifnature
  \bmhead{Author contributions}
S.A., R.G., M.K., F.C.G and A.K. contributed to the conception and design of the study. S.A. implemented and integrated the end-to-end pipeline, including key technical modules, and led training, validation, and manuscript drafting. R.G. developed the Dose Proposer component, and M.K. developed the deep learning dose calculation module. S.A., R.G., and M.D. contributed to data preparation. W.V., A.M., V.K., and S.B. provided clinical development, evaluation, and interpretation. F.C.G., A.K., and D.C. provided scientific leadership, oversight, and strategic guidance. The REQUITE Consortium provided the primary dataset. All authors reviewed and approved the final manuscript.
\fi

\newif\ifsupp
\supptrue

\ifsupp
	\renewcommand{\thefigure}{S\arabic{figure}}
	\renewcommand{\thetable}{S\arabic{table}}
	\newcounter{supnote}
	\setcounter{supnote}{0}
	\newcommand{\supsection}[1]{%
	  \refstepcounter{supnote}%
	  \section*{Supplementary Note~S\arabic{supnote}: #1}%
	  \addcontentsline{toc}{section}{Supplementary Note~S\arabic{supnote}: #1}%
	}
	\setcounter{figure}{0}
	\setcounter{table}{0}
	\clearpage
	\appendix
	\addcontentsline{toc}{section}{Supplementary Information}
	\supsection{Architecture and Implementation Details}
\label{sec:arch_and_implementation_details}

\subsection*{Structures and derivations}

The following table~\ref{tab:structures}  lists the structures, i.e.,  segmented organs and additional ``helper" structures derived from a mathematical formula.
These helper structures are used by RapidPlan / Eclipse optimization. The structures used by our AI pipelines are: PTV, bladder, rectum, and body. 

\begin{table}[ht]
\centering
\caption{Target and organ-at-risk (OAR) structures used in our Prostate RapidPlan.}
\begin{tabular}{@{}lll@{}}
\toprule
\textbf{Name} & \textbf{Type} & \textbf{Definition / Formula} \\
\midrule
CTV         & Target  & \( \text{Prostate} \cup \big((\text{Prostate}+3\,\text{mm}) \cap \text{Seminal Vesicles}\big) \) \\
PTV         & Target  & \( \text{CTV} + 3\,\text{mm} \) \\
PTV  $\cap$ Rectum  & Derived  &  \( \text{PTV} \cap \text{Rectum}  \)  \\
PTV Ring    & Derived & \( (\text{PTV}+20\,\text{mm}) - (\text{PTV}+5\,\text{mm}) \) \\
PTV--CTV    & Derived & \( \text{PTV} - \text{CTV} \) \\
50\% Ring   & Derived & \( (\text{PTV}+50\,\text{mm}) - (\text{PTV}+15\,\text{mm}) \) \\
Body Ring   & Derived & \( \text{Body} - (\text{Body}-35\,\text{mm}) \) \\
\midrule
Rectum        & OAR &  \\
Bowel\_Small  & OAR &  \\
Bowel\_Large  & OAR &  \\
Bladder       & OAR &  \\
Femur\_Head\_L & OAR &  \\
Femur\_Head\_R & OAR &  \\
Body          & Reference &  \\
\bottomrule
\end{tabular}
\label{tab:structures}
\end{table}

\subsection*{Data Augmentation}

We developed a Python script that applies different randomized transforms to the CT and contours in order to produce new pairs of CT and RT STRUCT files used to create additional plans in Eclipse.
The different transformations were: uniform scaling of the image size (in the range$[-20\%, +20\%]$), rotations in different axes (in the range  $[-5^\circ, +5^\circ]$), non-rigid deformation using the Continuous Piecewise-Affine based (CPAB) transformation ~\cite{freifeld2017transformations, detlefsen2018}, and random margins adjustments for the PTV and rectum contours in the range $[3\,\mathrm{mm}, 6\,\mathrm{mm}]$ and $[0\, \mathrm{mm}, 5\, \mathrm{mm}]$, respectively.
Each of those transforms was applied with a probability of 0.5, and 10 rounds of augmentation were performed, resulting in 11,180 additional plans to the original 1,277 plans, for a total of 12,457 plans.

\subsection*{Auto-Contouring (pre-processing)}
%
%

Our end-to-end pipeline takes the CT and contours as input. As the contours are derived from the CT.
The only pre-processing step to use our pipeline is to segment the organs at risk (OARs), the body,  and the planning target volume (PTV). Additional contours and helper structures are used by RapiPlan \cite{varian2024rapidplan, varian2024pelvisProstateSBRT} for the planning of our targets, but are not used during inference.
We used AI-Rad Companion Organs RT~\cite{siemens2024airad} as our automatic segmentation tool.
PyESAPI \cite{varian2024pyesapi} and Python were used for scripting and interaction with the Eclipse treatment planning system (TPS). A deeper description of our methodology is available in our previous publication~\cite{gao2025automating}.

In the following, we describe the modules of our end-to-end pipeline and use the same module names as in Fig.~\ref{fig:training_pipeline} of the main manuscript. We are using the following notation to describe each module as a function: $f_{\textsc{ModuleName}}(\cdot)$.

\subsection*{Dose proposer Network}

%

The first module of our end-to-end network is the dose proposer \cite{gao2023flexible}, a CNN with a 3D Res-UNet backbone. It processes four input channels: the CT volume and RT structures (Body, PTV, OARs). The two OARs contours, i.e. bladder and rectum, are merged into one channel using integer labels (1 for bladder, 2 for rectum, 3 for their overlap). 
CT Hounsfield Units values are clipped in the range [–900, +900] and rescaled to the range [0, 1]. 
We center the target dose around the isocenter and crop it along the inferior-superior direction to an extent of 64mm.
The dose proposer module can be interpreted as a function that predicts a dose distribution $D_{\mathrm{target}}$ conditionally on its input CT and structures $S$.
\begin{equation}
D_{\mathrm{target}} = f_{\mathrm{DoseProposer}}(\mathrm{CT}, S).
\end{equation}

The architecture of the network is depicted in Figure~\ref{fig:doseproposer}.
The network contains 17 millions parameters.
The network begins with a 3D convolutional (conv3D) stem (kernel size 3x3x3, stride 1, padding 1), followed by four residual blocks.
Each residual block contains a conv3D (kernel size 3x3x3, stride 1, padding 1), followed by an instance normalization and a ReLU activation, another conv3D and another instance normalization.
In parallel, the shortcut path employs a 1x1x1 convolution and an instance normalization. 
After that a 3D max pooling layer (kernel size 3, stride (2,2,2), padding 1) performs the downsampling step.
The number of channels increases from 4 at the input to 16 after the stem convolution, and then to 64, 128, and 256 at the first conv3Dlayer of the next three residual blocks.
The final residual block maintains 256 channels.
The decoder is using conv blocks containing a conv3D (kernel size 3, stride 1, padding 1), followed by an instance normalization and a ReLU activation.
The updampling step is implemented with trilinear interpolations.
The end of the network contains a 1x1x1 convolution layer to project the result to a single channel, followed by a Softplus activation function.

\begin{figure}[ht]
    \centering
    \includegraphics[width=\textwidth]{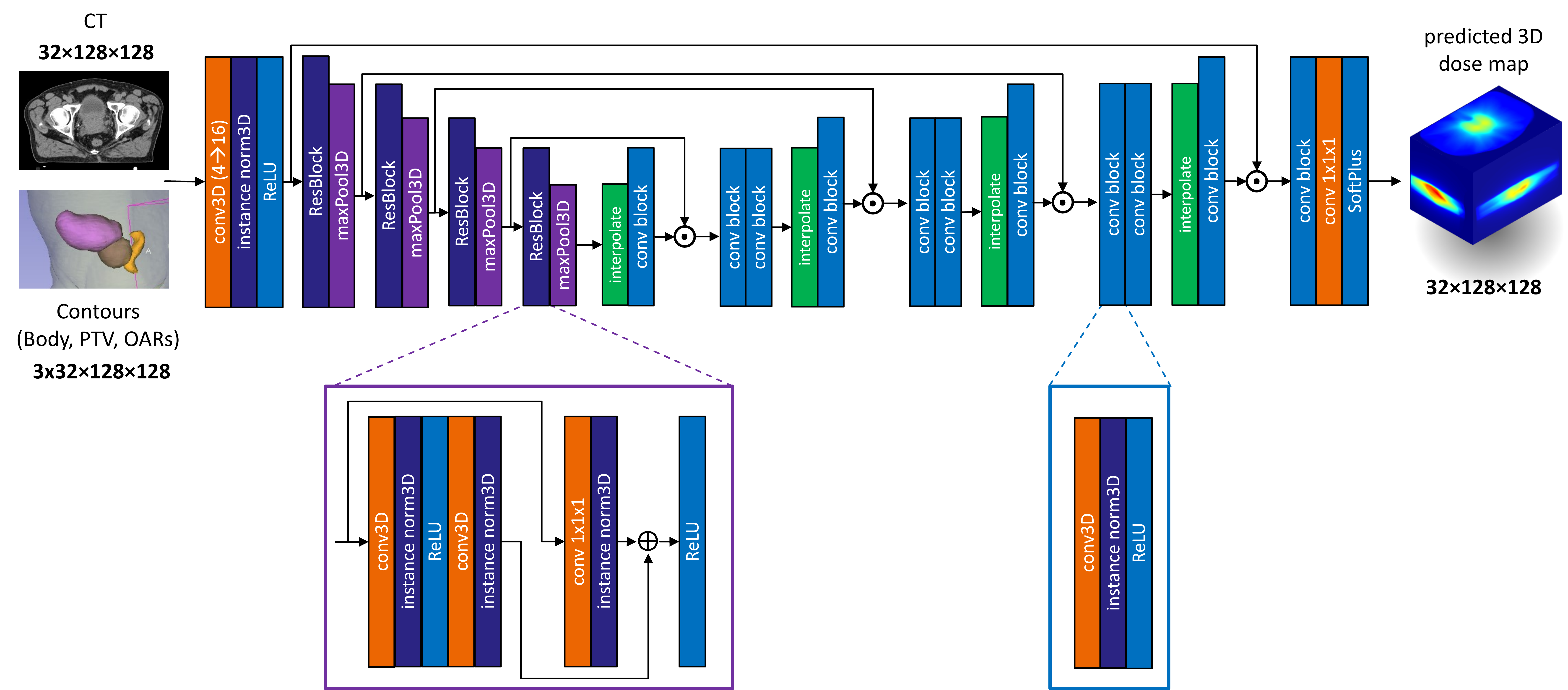}
    \caption{Architecture of the DoseProposer network.
    }
    \label{fig:doseproposer}
\end{figure}


\subsection*{Beam’s-Eye-View projections: BEV Proj}

%
%

The beam’s-eye-view (BEV) projection module, , depicted in Figure~\ref{fig:bevproj}, projects a 3D dose into the BEV corresponding to each control point of the VMAT plan, i.e., in the same reference geometry as the fluence maps. This geometric alignment between the inputs and outputs of the Bev2Fluence helps the network to perform its task efficiently.
The implementation of that projection was detailed in our previous article~\cite{arberet2025beam} on the Bev2Fluence network. We re-implemented this module, which has no trainable parameters, in PyTorch to have a differentiable module running efficiently on GPU.
Our BEV has a 5mm isotropic resolution and a 20 cm x 20 cm field-of-view.

We compute the BEV projection $\Pi_{\mathrm{BEVProj}}^{(cp)}(\cdot)$ on each control point $cp \in \{1,\dots,N_{\mathrm{cp}}\}$ and stack the result to create 3D volume:
\begin{equation}
B_{\mathrm{target}} = \left[\Pi_{\mathrm{BEVProj}}^{(cp)}\!\left(D_{\mathrm{target}}\right)\right]_{cp=1}^{N_{\mathrm{cp}}}.
\end{equation}

\begin{figure}[ht]
    \centering
    \includegraphics[width=0.5\textwidth]{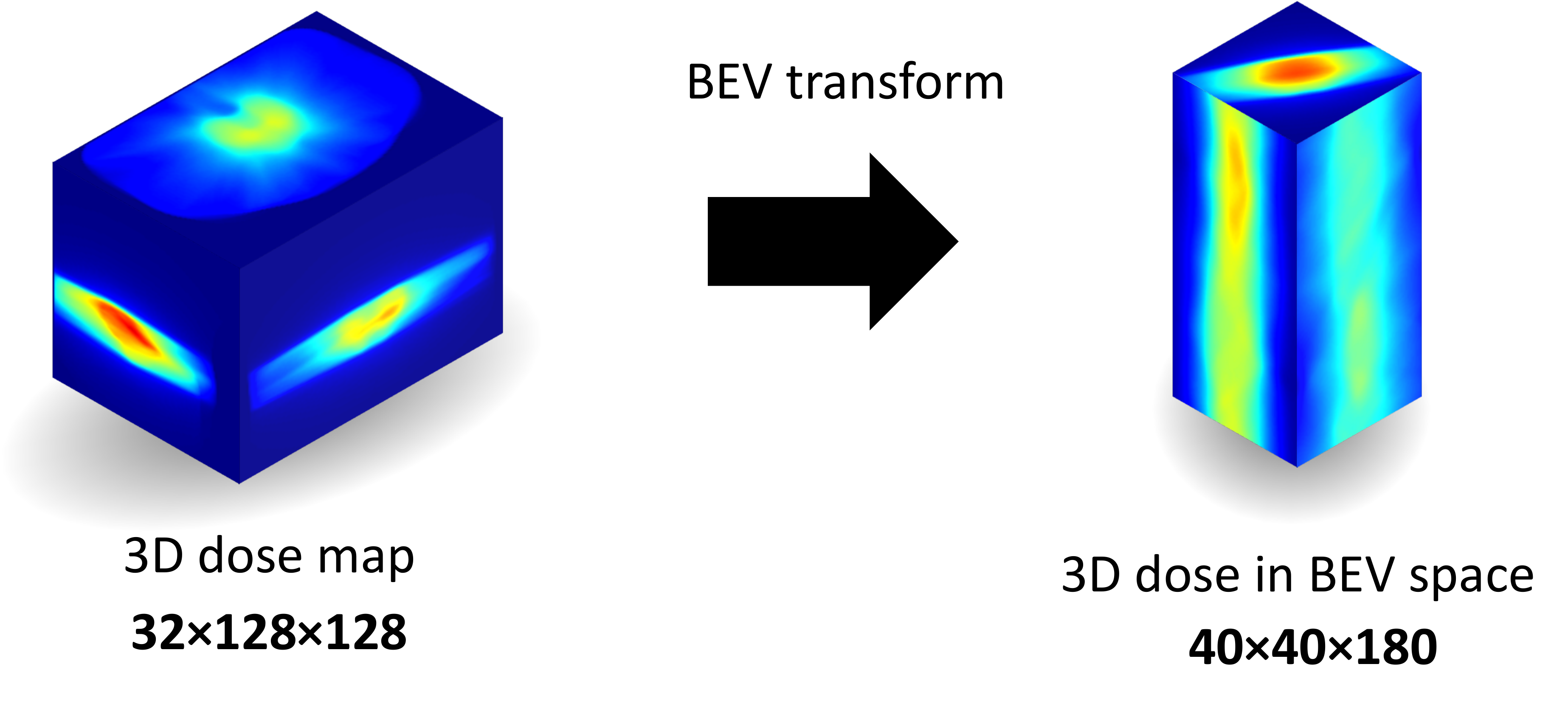}
    \caption{Beam's eye view (BEV) transform.
    }
    \label{fig:bevproj}
\end{figure}

\subsection*{Bev2Fluence Network}

%

The Bev2Fluence network \cite{arberet2025beam}, depicted in Figure~\ref{fig:bev2fluence}, is a 40.8 M parameters 3D CNN, with a MedNeXT backbone architecture \cite{roy2023mednext}, that predicts the 180 VMAT fluence maps from the 180 BEV input projections.

The 3D MedNeXt backbone \cite{roy2023mednext} is an encoder-decoder structure with four resolution levels.
Each level contains MedNeXt blocks, followed by downsampling layers implemented via strided depthwise convolutions.
The decoder mirrors the encoder structure with transposed convolutions and skip connections.
Each MedNeXt block contains internal channel expansion including in the bottleneck. 
Each convolution block uses group normalization and GELU activation functions.

The Bev2Fluence network \cite{arberet2025beam}, based on the MedNeXt backbone, first project the input BEV dose volume to 64 channels via a 1x1x1 stem convolution.
The number of channels then increases from 64, to 128, 256, and 512 at each downsampling step.
The bottleneck operates at 1024 channels with internal expansion to 4096 channels.
The decoder mirrors the encoder structure and a final 1x1x1 transposed convolution project the 64 channel volume into a single channel output, followed by an abs activation layer to prevent negative coefficients from being inputed into the DL dose module. 
We refer the reader to our previous work \cite{arberet2025beam} and the original MedNeXT \cite{roy2023mednext} article, which contain additional details about this network architecture.
The functional description of the Bev2Fluence network is as follows:
\begin{equation}
F^{(0)} = f_{\mathrm{Bev2Fluence}}\!\left(B_{\mathrm{target}}\right),
\end{equation}
where $F^{(0)}=\left[F^{(0)}_{cp}\right]_{cp=1}^{N_{\mathrm{cp}}}$ is the stack of predicted fluence maps.

\begin{figure}[ht]
    \centering
    \includegraphics[width=\textwidth]{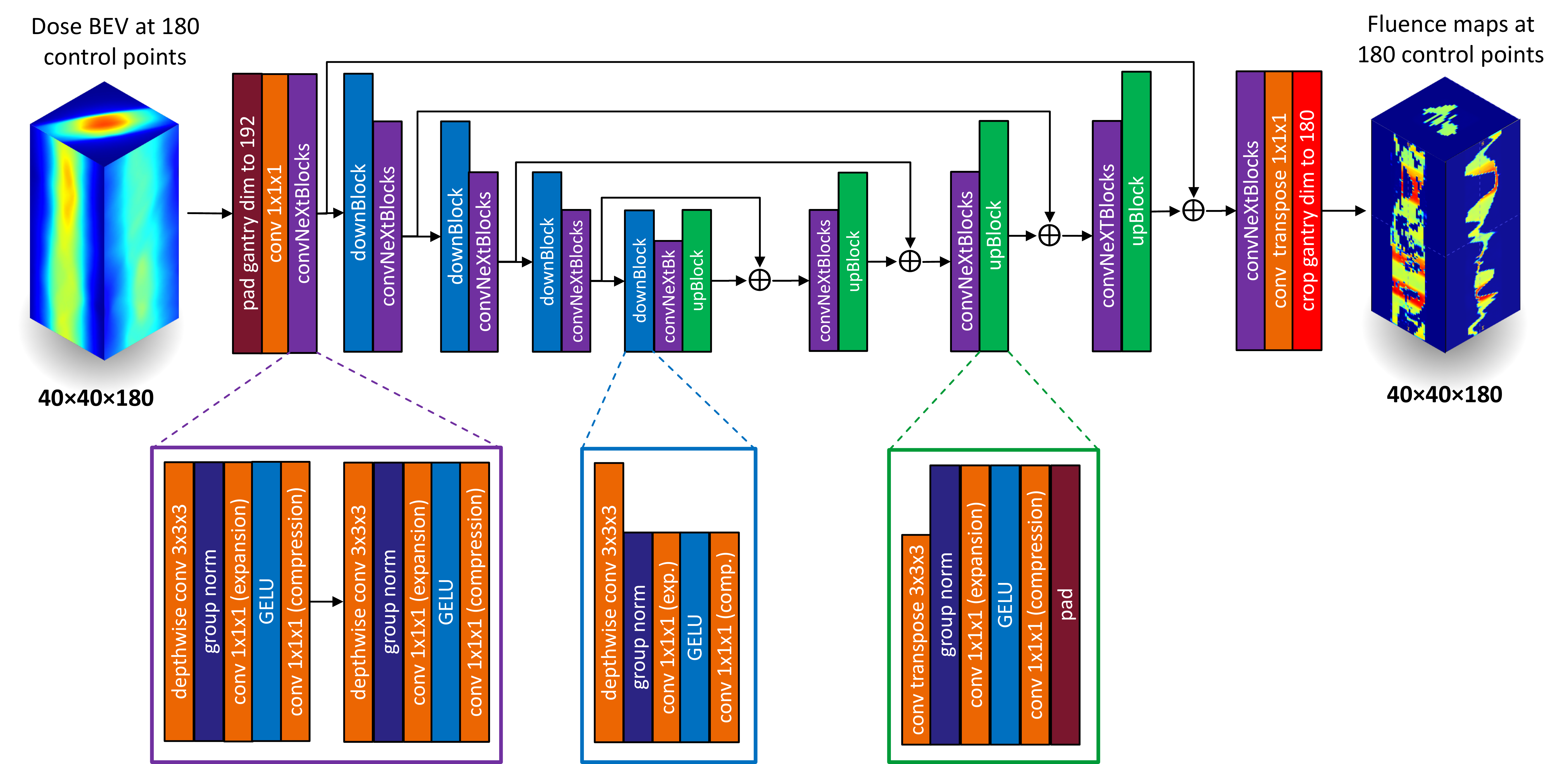}
    \caption{Architecture of the Bev2Fluence network.
    }
    \label{fig:bev2fluence}
\end{figure}

\subsection*{Dose Computation Network}

The dose computation network uses a physics-informed approach~\cite{kraus2025singleshot} that predicts the 3D dose volume from a stack of input fluence maps and their physical descriptions (gantry angles, collimator angle, field-of-view, etc.).

We use this module as a differentiable evaluator to obtain the dose delivered by the predicted fluence maps:
\begin{equation}
D^{(0)} = f_{\mathrm{DLDoseComputation}}\!\left(\mathrm{CT}, F^{(0)}\right).
\end{equation}

It adopts a two-stage design, depicted in Figure~\ref{fig:dl_dose_two_stage}, where first un-collided 3D fluence is computed and accumulated as spherical harmonics coefficients per voxel, and then in stage 2, these coefficients together with the CT are processed throught a MedNeXt image-to-image network to compute the final 3D dose map.

This network contains 2.8 M parameters and begins with a 1x1x1 stem convolution layer that project the 26-channel input volume into a 16 channel feature map.
The number of feature channels increases at each downsampling operation from 16 to 32, 64, 128, and 256. The bottleneck operates at 256 channels with internal expansion to 1024 channels.
The decoder mirros the encoder structure and ends with a 1x1x1 convolution projecting the 16-channel feature maps to a single-channel output, followed by a leaky-ReLU activation (negative slope 1e-2).

\begin{figure}[t]
\centering

\includegraphics[width=0.65\textwidth]{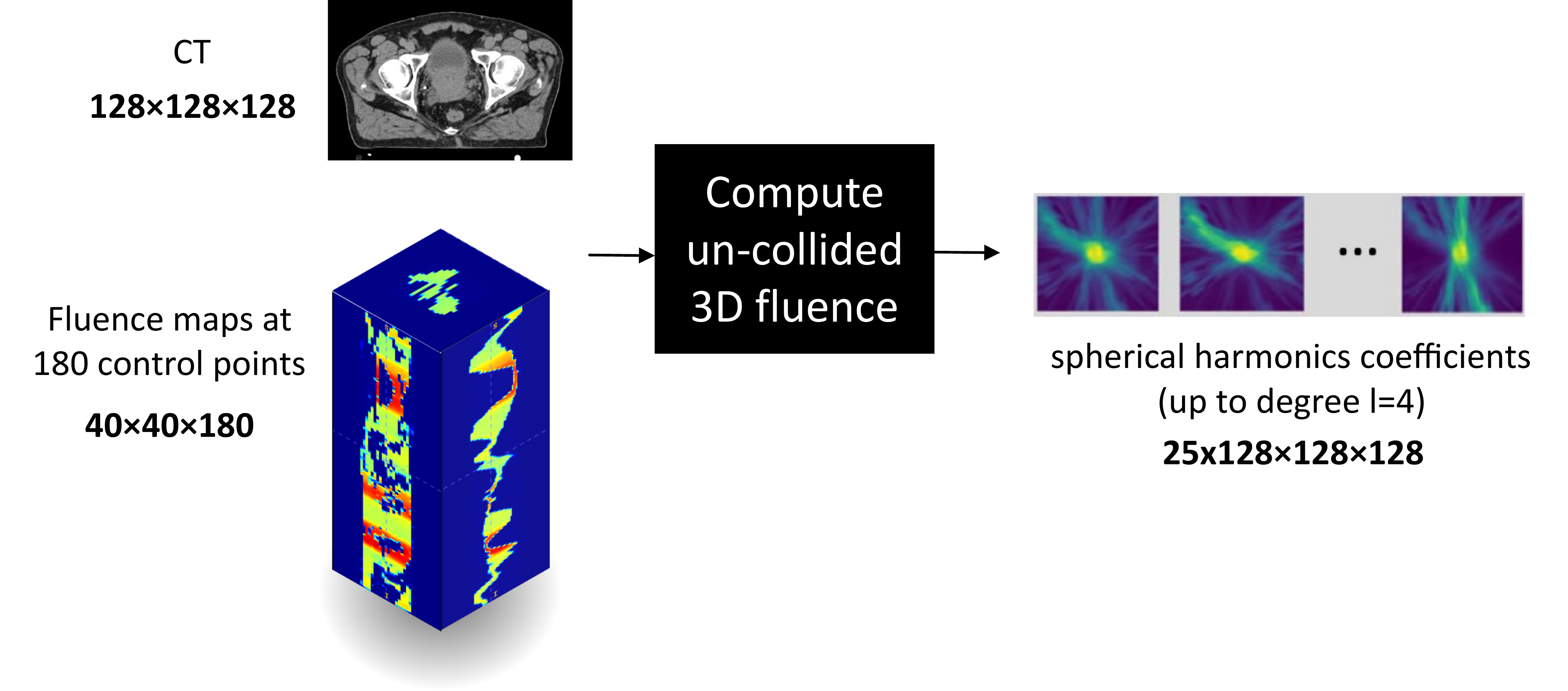}
\\[6pt]
\small \textbf{a} Stage 1: Uncollided fluence computation

\vspace{12pt}

\includegraphics[width=\textwidth]{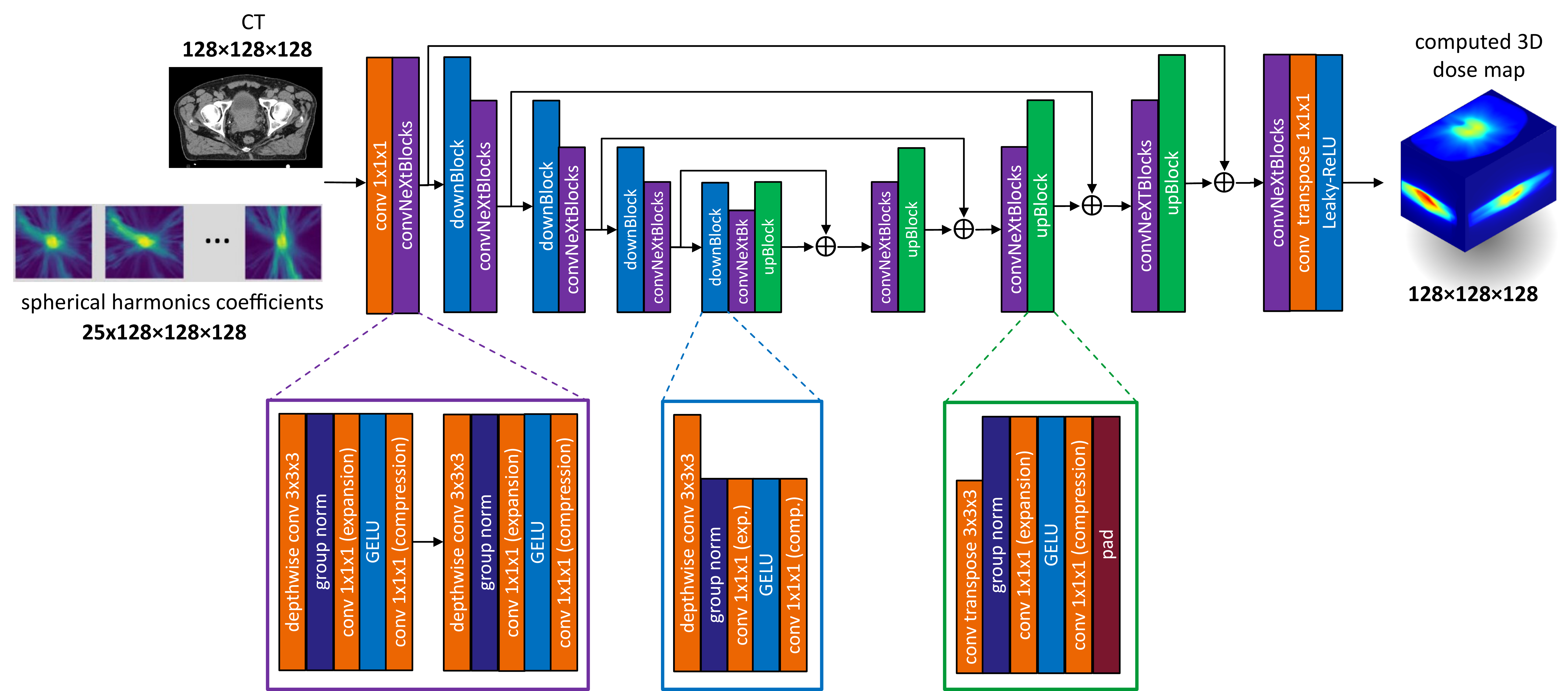}
\\[6pt]
\small \textbf{b} Stage 2: Final dose computation

\caption{\textbf{Two-stage deep learning dose computation.}
\textbf{a,} The first stage predicts the voxel-wise spherical harmonics coefficients from the planning CT and fluence maps. 
\textbf{b,} The second computes the 3D dose map from the CT and spherical harmonics coefficients using a MedNeXT network architecture.}
\label{fig:dl_dose_two_stage}
\end{figure}

\subsection*{Dose Error Correction: (\errModule)}


The purpose of the dose error correction module (\errModule), which has no trainable parameters, is to generate a dose error signal, based on the input dose map and some dose metrics (e.g. target inhomogeneity). This dose error signal is first used in the pipeline to correct the initial set of fluence maps (at both training and inference), as well as in the loss to reduce that error based on the training population data (at training only). 
Without it, predicting 180 VMAT fluence maps in a single shot is extremely challenging. 
In the following, we denote the voxel-wise error map as $E^{(0)}$.

\paragraph{Error map as the gradient of a dose objective.}
We define the error map in general terms as the gradient with respect to the 3D dose map of a differentiable objective function $J(D)$:
\begin{equation}
E^{(0)} \;=\; f_{\mathrm{Err}}\!\left(D^{(0)}, \cdot\right) \;\approx\; \nabla_{D}\, J\!\left(D^{(0)}\right),
\label{eq:Err_as_grad}
\end{equation}
where $D^{(0)}\in\mathbb{R}^{|\Omega|}$ is the dose computed from the initial fluence maps, and $|\Omega|$ is the number of voxels in the dose grid.

In its generic form, the dose objective $J(D)$ can be written as a sum over the structures $s$ (PTV, bladder, rectum, etc), and for each structure, a sum over structure-specific objectives $i$ as: 
\begin{equation}
J(D) \;=\; \sum_{s\in\mathcal{S}} \sum_{i\in\Omega} M_s(i)\,\phi_s\!\bigl(D_i - R_{s,i}\bigr),
\label{eq:J_general}
\end{equation}
where $M_s(i)\ge 0$ is  a mask (over a grid) or weight (scalar) for structure $s$. 
$R_s\in\mathbb{R}^{|\Omega|}$ is a clinical goal or reference, and $\phi_s(\cdot)$ is a function that encodes the penalty (e.g. L2 norm, ReLU).


\paragraph{PTV term: asymmetric hot-spot suppression.}

In our study, we designed the PTV term in an asymmetric manner, in order to penalize hot spots (overdose) more strongly than cold spots (underdose).
Our PTV homogeneity term objective can be written as:
\begin{equation}
J_{\mathrm{PTV}}(D)
\;=\;
\sum_{i\in\Omega} M_{\mathrm{PTV}}(i)
\left(
\frac{\lambda_{+}}{2}\,\mathrm{ReLU}(x_i)^2
+
\frac{\lambda_{-}}{2}\,\mathrm{ReLU}(-x_i)^2
\right),
\label{eq:J_ptv_final}
\end{equation}
where $M_{\mathrm{PTV}}$ denotes the PTV mask, $D_{\mathrm{Rx}}$ the prescribed dose (scalar), and 
$
x_i \;=\; D_i - D_{\mathrm{Rx},i}.
$

Deriving the (voxel-wise) gradient of Eq~\eqref{eq:J_ptv_final} with respect to the input dose $D$ leads to the definition of the corrected signal we used in our study for the PTV:
\begin{equation}
\begin{aligned}
\left[E^{(0)}_{\mathrm{PTV}}\right]_i
&=
\left[\nabla_D J_{\mathrm{PTV}}(D)\right]_i\Big|_{D=D^{(0)}} \\[4pt]
&=
M_{\mathrm{PTV}}(i)\Bigl(
\lambda_{+}\,\mathrm{ReLU}\!\left(D^{(0)}_i-D_{\mathrm{Rx},i}\right)
-
\lambda_{-}\,\mathrm{ReLU}\!\left(D_{\mathrm{Rx},i}-D^{(0)}_i\right)
\Bigr).
\end{aligned}
\label{eq:E_ptv_signed}
\end{equation}

In order to penalize the hot spots more than the cold spots, we used $\lambda_{+}>\lambda_{-}>0$. 


\paragraph{Optional OAR term: user-controlled dose suppression.}

Rather than using a complex set of DVH-based clinical goals, we opted for providing the user with a simple way to control OAR sparing.
We define a single scalar suppression factor $s_o$ which can steer the baseline plan, i.e., the plan predicted by our pipeline when no OAR sparing is used.

For a given OAR $o$ and its suppression factor $s_o$, we define the reference field as:
\begin{equation}
R_{o,i} \;=\; (1-s_o)\,D^{(0)}_i,
\label{eq:Ro_final}
\end{equation}
where $D^{(0)}$ is the dose predicted at stage 0.

We then use the squared-hinge objective function:
\begin{equation}
J_{o}(D)
\;=\;
\frac{1}{2}\sum_{i\in\Omega} M_o(i)\,\Bigl(\max\!\bigl(D_i - R_{o,i},\,0\bigr)\Bigr)^2.
\label{eq:J_oar_final}
\end{equation}
which gradient, evaluated at $D^{(0)}$ produces the OAR dose correction term used in our implementation:
\begin{equation}
\left[E^{(0)}_{\mathrm{OAR},o}\right]_i
\;=\;
\left[\nabla_D J_{o}(D)\right]_i\Big|_{D=D^{(0)}}
\;=\;
M_o(i)\,\max\!\bigl(D^{(0)}_i - R_{o,i},\,0\bigr).
\label{eq:E_oar_final}
\end{equation}
We can note that when $s_o=0$, the OAR penalty is disabled ($R_o=D^{(0)}$), while increasing $s_o>0$ decreases the reference $R_o$ and increases the dose penalty for that OAR.


\paragraph{Total error map and BEV projection.}
The full error map is obtained by combining all the error terms (PTV and optional OARs):
\begin{equation}
E^{(0)} \;=\; E^{(0)}_{\mathrm{PTV}} \;+\; \sum_{o\in\mathcal{O}} E^{(0)}_{\mathrm{OAR},o},
\label{eq:E_total_final}
\end{equation}
and is then projected into the BEV of each control point:
\begin{equation}
B_{\mathrm{err}}^{(0)} = \left[\Pi_{\mathrm{BEVProj}}^{(cp)}\!\left(E^{(0)}\right)\right]_{cp=1}^{N_{\mathrm{cp}}}.
\label{eq:Bev_err_final}
\end{equation}
This projected tensor is then used as the input of the Bev2Fluence correction network, which completes the dose feedback loop.
The Bev2Fluence correction network, depicted in Figure~\ref{fig:bev2FluenceCorrection}, has the same architecture as the Bev2Fluence network, except that it accepts two input volumes instead of one.

\begin{figure}[ht]
    \centering
    \includegraphics[width=\textwidth]{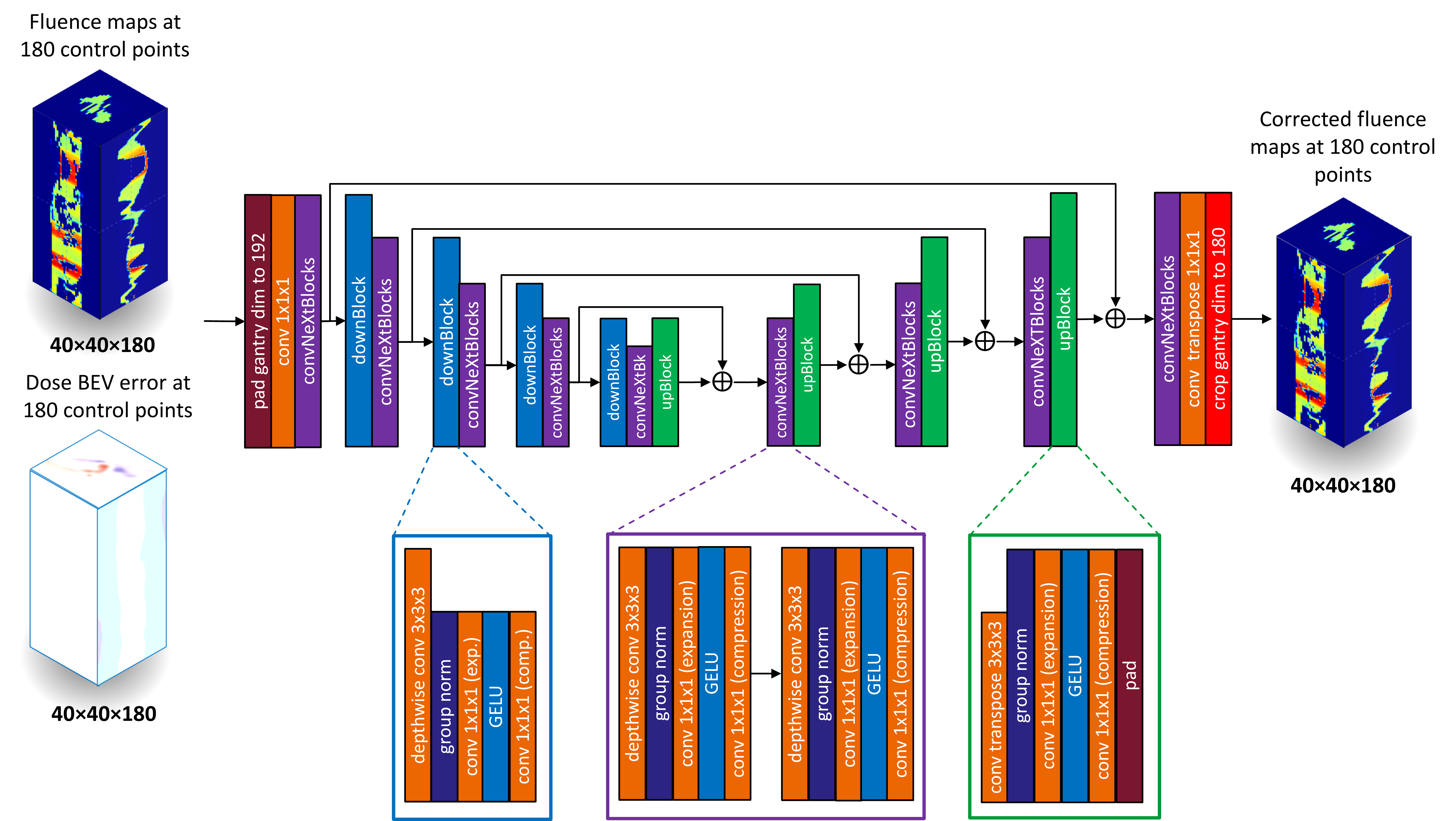}
    \caption{Architecture of Bev2Fluence correction network.
    }
    \label{fig:bev2FluenceCorrection}
\end{figure}

\subsection*{Fluence Correction Network}


The fluence correction network adopts the same architecture as the Bev2Fluence network, except that its number of input channels is increased to receive the initial fluence maps in addition to the BEV dose error projection. It outputs the corrected fluence maps $F^{(1)}$:
\begin{equation}
F^{(1)} = f_{\mathrm{Bev2FluenceCorrection}}\!\left(F^{(0)}, B_{\mathrm{err}}^{(0)}\right).
\end{equation}
This module could be used for multiple iterations $k$: $$F^{(k+1)} = f_{\mathrm{Bev2FluenceCorrection}}\!\left(F^{(k)}, B_{\mathrm{err}}^{(k)}\right),$$ with either shared or separated weights (unrolling) in order to improve further the dose performance of the pipeline at the expense of added computation.
In our study, we used only one correction loop in order to preserve computational efficiency and minimize memory usage.

\subsection*{Adversarial Fluence Shaping: Implementation Details}
\label{supp-sec:adv-fluence}


Our adversarial training approach, applied to our fluence maps, consists of a discriminator trained to classify whether a fluence map has been generated by our end-to-end pipeline or from a VMAT plan optimized in Eclipse.
Our discriminator implementation is similar to the one used in StyleGAN2~\cite{karras2020analyzing, lucidrains_stylegan2_pytorch}.
The architecture of the network is depicted in Figure~\ref{fig:discriminator}.
It consists of a sequence of downsampling blocks that progressively downsample the input image while increasing the feature dimension.
The network has 22.7 M parameters and receives a single 64 x 64 2D fluence map as input which was zero-padded from its original 40 x 40 size. The 180 fluence maps are processed in parallel by staking them on the batch dimension.
The first block project the 2D input to 64 feature channels using a 1 x 1 convolution with stride 2, followed by two 3 x 3 convolutions with leaky-ReLU activations (negative slope 0.2) as well as a downsampling operation performed via a combination of a low-pass filtering (Blur) and a strided 3 x 3 convolution.  The downsampling block ends with a residual connection. 
This downsampling block is repeated five times with feature resolutions decreasing from 64 to 2 by factors of 2 while the number of feature channels are increasing from 64 to 512 by a factor of 2 (the final convolution block maintains the 512 channels).
After that, the network ends with a final 3 x 3 convolution, a flattening layer reshaping the features to a single vector of size 2048, and a linear (fully-connected) layer to produce the scalar logit output.

\begin{figure}[ht]
    \centering
    \includegraphics[width=0.5\textwidth]{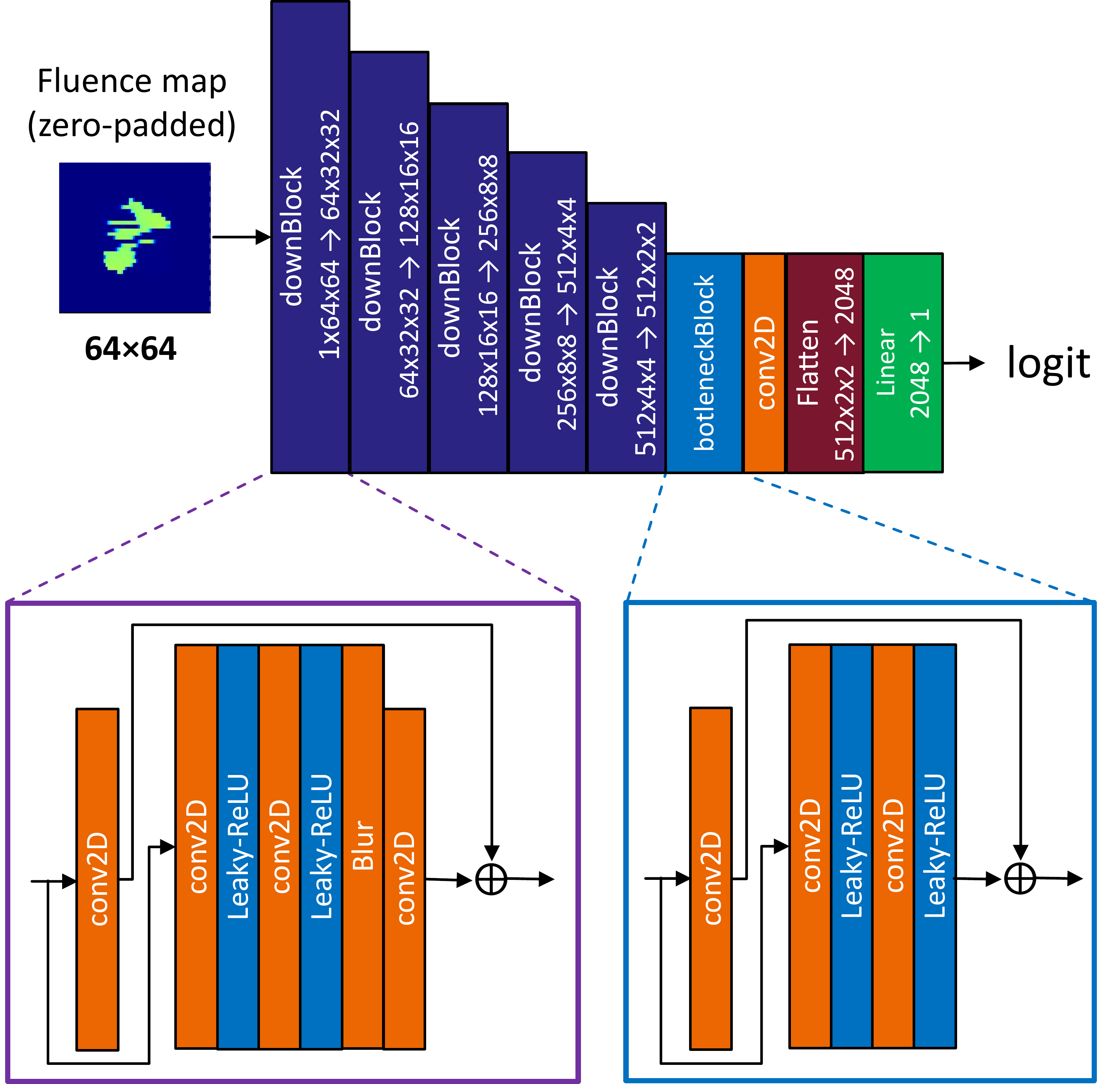}
    \caption{Architecture of fluence map discriminator network.
    }
    \label{fig:discriminator}
\end{figure}

The discriminator was trained using a Hinge loss with a R1 gradient penalty
 on real (target fluence maps) samples in order to encourage local smoothness of the discriminator function. 
I other words, if $D(\cdot)$ denote the discriminator output (without sigmoid),  and $G(\cdot)$ the generator (the end-to-end pipeline in our case), the discriminator can be written as:
\begin{equation}
\label{eq:discr_hinge_loss}
\mathcal{L}_D^{\mathrm{total}} =
\mathcal{L}_D
+
\frac{\gamma}{2}
\mathbb{E}_{x \sim p_{\mathrm{data}}}
\left[
\left\| \nabla_x D(x) \right\|_2^2
\right], 
\end{equation}
where the second term is the R1 penalty, and the first term $\mathcal{L}_D $ is the adversarial Hinge loss defined by:
\begin{equation}
\mathcal{L}_D =
\mathbb{E}_{x \sim p_{\mathrm{data}}}
\left[
\max(0,\, 1 - D(x))
\right]
+
\mathbb{E}_{z \sim p(z)}
\left[
\max(0,\, 1 + D(G(z)))
\right].
\end{equation}

The generator loss, i.e., the loss term used to regularize the fluence maps generated by our end-to-end pipeline, was a generator Hinge loss, which is the negative of the average discriminator score of ``fake" samples (i.e. generated by our end-to-end pipeline).
\begin{equation}
\label{eq:gen_hinge_loss}
\mathcal{L}_G =
- \mathbb{E}_{z \sim p(z)} \left[ D(G(z)) \right].
\end{equation}

The effects of our adversarial training on the resulting fluence maps produced by our pipeline are illustrated in Fig.~\ref{fig:adv-fluence-mosaic}.

\begin{figure}[htbp]
    \centering
    \includegraphics[width=\textwidth]{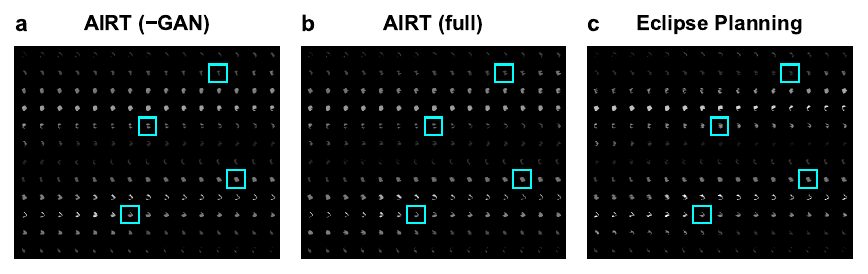}
    \vspace{1em}
    \includegraphics[width=0.85\textwidth]{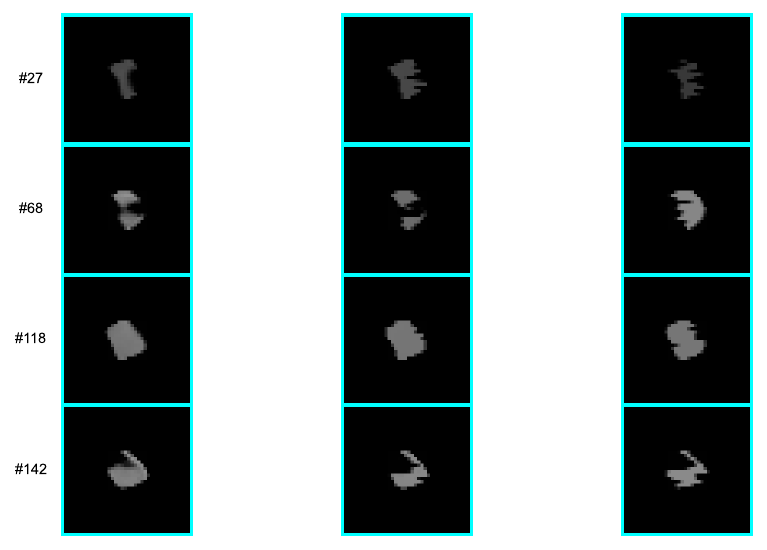}
    \caption{
        \textbf{Adversarial fluence shaping results.}
        (\textbf{Top}) Mosaic view comparing fluence maps generated by (\textbf{a}) \AIRT{}~(-GAN, i.e. without adversarial loss), (\textbf{b}) \AIRT{}~(full, i.e. with adversarial loss), and (\textbf{c}) Eclipse clinical plans.
        (\textbf{Bottom}) Zoom-in of representative fluence maps highlighing the improved realism achieved with adversarial training. Both views illustrate the discriminator’s impact in enforcing clinically plausible, approximate two-level structure in generated fluence maps.
    }
    \label{fig:adv-fluence-mosaic}
\end{figure}

\supsection{Training Strategy and Optimization Details}
\label{sec:supp_training}

%
%
%
%
%

%

\subsection*{Stage 1: Full pipeline training without adversarial loss}

We first train the pipeline without the adversarial loss, as illustrated in the top panel of Fig.~\ref{fig:training_pipeline} of the main manuscript.
In both training stages, we use the ADAM optimizer from PyTorch with a fixed learning rate of 1e-4 and a batch size of 1.

We have the following losses:
\begin{itemize}
\item \textit{Dose Proposer loss}: To train the Dose Proposer network to generate plausible 3D dose distributions, we minimized the L1 norm loss between the network’s predicted dose and the target dose, where the target was computed using our deep-learning dose calculation applied to the Eclipse-optimized plans of our training dataset.
\item \textit{Fluence maps loss stage 1}:  To train the Bev2Fluence network to generate plausible fluence maps, we minimized the L1 norm loss between this network’s predicted fluence maps and the target fluence maps, i.e. the fluence maps of the Eclipse-optimized plans of our training dataset.
\item \textit{Dose loss stage 1}: The fluence maps predicted by the Bev2Fluence module are converted to a 3D dose map using our DL dose computation and a L1 loss is computed between this predicted dose map and the target dose map.
\item \textit{Dose error loss stage 1}: 
The \textit{Dose error loss stage 1} is a light regularizer (we used it with a very small weight) that encourages PTV homogeneity.
Notably, this loss is only meaningful when the \errModule{}  metric does not include variable OAR sparing controls, since at this stage, the network cannot yet respond to such commands: it needs first to predict a baseline for the stage 1. The ``customization'' of that baseline happens in stage 2 of training, using the Bev2Fluence correction network.
\item \textit{Fluence map loss}: To train the Bev2Fluence Correction Network which predicts the final fluence maps, we still apply a L1 loss between these predicted fluence maps and the target fluence maps, similarly as the \textit{Fluence maps loss stage 1}.
\item \textit{Dose error loss}: Finally, to train the Bev2Fluence Correction Network to effectively utilize its input 3D dose error map as a correction signal, we recompute a new 3D dose error map using the DL dose and \errModule{} module, and apply a loss function that minimizes the L1 norm of that final error.
\end{itemize}
We used the following weights for each loss of stage 1 training:
\textit{dose proposer loss}: 0.3, \textit{fluence maps loss stage 1}: 0.1, \textit{dose loss stage 1}: 0.1, \textit{dose error loss stage 1}: 0.0003, \textit{fluence maps loss}: 0.4, \textit{dose error loss}: 0.001

\subsection*{Stage 2: Fluence correction network training with adversarial loss}

The second stage of training has two main objectives:
(1) Adding the adversarial loss to enforce the fluence maps predicted by the Bev2Fluence correction module to be close to the manifold of VMAT optimized fluence maps, and thereby to facilitate leaf-sequencability.
(2) Training the Bev2Fluence correction module to adjust its behavior to variable user-specified OAR sparing controls, in addition to maintaining PTV homogeneity.
To implement these two objectives, we (1) trained a discriminator with the loss function Eq. \eqref{eq:discr_hinge_loss} and added the term Eq. \eqref{eq:gen_hinge_loss} in the generator (the pipeline) loss, as detailed in the subsection on Adversarial Fluence shaping in section~\ref{sec:arch_and_implementation_details}, and (2) randomized the parameters of the \errModule{} module for each training batch to cover the variability of dose objectives (OAR sparing factors) use cases.

During this second stage of training, we froze all the network components before the Bev2Fluence Correction module, as these components, which were trained in stage 1 are expected to predict an initial baseline fluence maps, not influenced by the variability of dose objectives. Unfreezing them would create contradictory backpropagation gradients, making the training unstable.
Another advantage of freezing these components is that training is significantly faster, as it eliminates the need to backpropagate to all these modules, particularly the DL dose computation module embedded in the pipeline, which is the most computationally intensive module.

We used the following weights for each loss of stage 2 training:
For the generator (pipeline) part:
\textit{fluence maps loss}: 0.2 
, \textit{generator hinge loss}: 0.04, 
\textit{dose error loss}: 0.001, 
and for the discriminator part, we used the Hinge loss function of Eq. \eqref{eq:discr_hinge_loss}, with $\gamma=0.0025$.


\supsection{Dosimetric Performance}
\label{sec:dose_performance}

In Table \ref{tab:comparison_all_engines} are depicted the dosimetric performance of the \AIRT method and the RapidPlan Eclipse evaluated with four different dose engines: DL dose, LTBS, Eclipse AAA and Eclipse AcurosXB.

\begin{table}[ht]
\centering
\caption{Comparison of Eclipse and \AIRT{} planning for intact prostate VMAT cases across 4 dose calculation engines. The homogeneity index (HI) is defined as $(D_{2\%} - D_{98\%}) / D_{50\%}$ following the ICRU convention (lower is better). All metrics are reported as mean $\pm$ standard deviation over the validation dataset. P-values are from the Wilcoxon signed-rank test.}
\label{tab:comparison_all_engines}
\begin{tabular}{llccc}
\toprule
\multicolumn{5}{c}{\textbf{DL dose engine}} \\
\midrule
\textbf{Structure} & \textbf{Metric} & \textbf{Eclipse Plans} & \textbf{AIRT Plans} & \textbf{p-value} \\
\midrule
\multirow{2}{*}{PTV} & HI & 0.16 $\pm$ 0.03 & 0.11 $\pm$ 0.02 & $<$ 0.001 \\
 & D98 (Gy) & 38.4 $\pm$ 0.8 & 38.8 $\pm$ 0.4 & $<$ 0.001 \\
\midrule
\multirow{3}{*}{Bladder} & Dmean (Gy) & 6.7 $\pm$ 3.3 & 6.9 $\pm$ 3.5 & 0.01 \\
 & D50 (Gy) & 2.9 $\pm$ 2.5 & 3.0 $\pm$ 2.7 & 0.29 \\
 & D2 (Gy) & 34.8 $\pm$ 6.8 & 35.1 $\pm$ 6.6 & 0.07 \\
\midrule
\multirow{3}{*}{Rectum} & Dmean (Gy) & 5.5 $\pm$ 1.4 & 5.3 $\pm$ 1.4 & 0.002 \\
 & D50 (Gy) & 2.7 $\pm$ 1.2 & 2.5 $\pm$ 1.1 & $<$ 0.001 \\
 & D2 (Gy) & 30.1 $\pm$ 5.5 & 30.5 $\pm$ 5.8 & 0.04 \\
\midrule
\multicolumn{5}{c}{\textbf{LTBS dose engine}} \\
\midrule
\textbf{Structure} & \textbf{Metric} & \textbf{Eclipse Plans} & \textbf{AIRT Plans} & \textbf{p-value} \\
\midrule
\multirow{2}{*}{PTV} & HI & 0.19 $\pm$ 0.03 & 0.15 $\pm$ 0.03 & $<$ 0.001 \\
 & D98 (Gy) & 37.8 $\pm$ 0.9 & 38.3 $\pm$ 0.6 & $<$ 0.001 \\
\midrule
\multirow{3}{*}{Bladder} & Dmean (Gy) & 6.8 $\pm$ 3.4 & 6.9 $\pm$ 3.5 & 0.05 \\
 & D50 (Gy) & 2.8 $\pm$ 2.3 & 2.8 $\pm$ 2.5 & 0.66 \\
 & D2 (Gy) & 36.3 $\pm$ 7.5 & 36.5 $\pm$ 7.3 & 0.50 \\
\midrule
\multirow{3}{*}{Rectum} & Dmean (Gy) & 5.1 $\pm$ 1.3 & 4.9 $\pm$ 1.3 & 0.002 \\
 & D50 (Gy) & 2.5 $\pm$ 1.0 & 2.2 $\pm$ 0.9 & $<$ 0.001 \\
 & D2 (Gy) & 30.2 $\pm$ 6.0 & 30.5 $\pm$ 6.4 & 0.03 \\
\midrule
\multicolumn{5}{c}{\textbf{Eclipse AcurosXB engine}} \\
\midrule
\textbf{Structure} & \textbf{Metric} & \textbf{Eclipse Plans} & \textbf{AIRT Plans} & \textbf{p-value} \\
\midrule
\multirow{2}{*}{PTV} & HI & 0.10 $\pm$ 0.01 & 0.10 $\pm$ 0.01 & 0.03 \\
 & D98 (Gy) & 39.3 $\pm$ 0.2 & 39.3 $\pm$ 0.2 & 0.36 \\
\midrule
\multirow{3}{*}{Bladder} & Dmean (Gy) & 7.3 $\pm$ 3.5 & 7.7 $\pm$ 3.8 & $<$ 0.001 \\
 & D50 (Gy) & 3.3 $\pm$ 2.7 & 3.6 $\pm$ 3.0 & $<$ 0.001 \\
 & D2 (Gy) & 35.8 $\pm$ 6.0 & 37.0 $\pm$ 5.9 & $<$ 0.001 \\
\midrule
\multirow{3}{*}{Rectum} & Dmean (Gy) & 5.4 $\pm$ 1.4 & 5.5 $\pm$ 1.4 & 0.66 \\
 & D50 (Gy) & 2.7 $\pm$ 1.0 & 2.6 $\pm$ 1.0 & 0.003 \\
 & D2 (Gy) & 31.4 $\pm$ 5.6 & 32.5 $\pm$ 6.0 & $<$ 0.001 \\
\midrule
\multicolumn{5}{c}{\textbf{Eclipse AAA engine}} \\
\midrule
\textbf{Structure} & \textbf{Metric} & \textbf{Eclipse Plans} & \textbf{AIRT Plans} & \textbf{p-value} \\
\midrule
\multirow{2}{*}{PTV} & HI & 0.10 $\pm$ 0.01 & 0.10 $\pm$ 0.01 & 0.28 \\
 & D98 (Gy) & 39.4 $\pm$ 0.2 & 39.3 $\pm$ 0.2 & 0.12 \\
\midrule
\multirow{3}{*}{Bladder} & Dmean (Gy) & 7.6 $\pm$ 3.6 & 8.0 $\pm$ 3.9 & $<$ 0.001 \\
 & D50 (Gy) & 3.6 $\pm$ 2.7 & 3.9 $\pm$ 3.1 & $<$ 0.001 \\
 & D2 (Gy) & 36.1 $\pm$ 6.0 & 37.2 $\pm$ 5.9 & $<$ 0.001 \\
\midrule
\multirow{3}{*}{Rectum} & Dmean (Gy) & 5.7 $\pm$ 1.4 & 5.7 $\pm$ 1.5 & 0.61 \\
 & D50 (Gy) & 3.0 $\pm$ 1.1 & 2.8 $\pm$ 1.1 & 0.003 \\
 & D2 (Gy) & 31.5 $\pm$ 5.7 & 32.6 $\pm$ 6.0 & $<$ 0.001 \\
\bottomrule
\end{tabular}
\end{table}

\supsection{Statistical Validation and Non-Inferiority}
\label{sec:supp_noninferiority}


In order to assess statistical equivalence between our end-to-end AI plans and the Eclipse plans, we conducted a non-inferiority test on the main DVH metrics using the Eclipse AcurosXB dose engine for the evaluation.
For each metric, we used a statistical significance of p $<$ 0.05 and a clinical margin of 0.01 for the homogeneity index metrics, and 1.5 Gy for the organ at risk (OAR) dose metrics.
As illustrated in Table~\ref{tab:supp_noninferiority}, AI planning met the non-inferiority criteria on all the tested metrics.

\begin{table}[ht]
\centering
\caption{Non-inferiority test results comparing our AI planning to Eclipse planning on DVH metrics evaluated using Eclipse AcurosXB dose engine.
A metric is considered non-inferior if the one-sided non-inferiority test is statistically significant (p $<$ 0.05), and the difference does not exceed the predefined margin.
Here, the mean difference is reported as the AI metric minus the Eclipse one.}
\begin{tabular}{lrrrrr}
\hline
\textbf{Metric} & \textbf{Mean Diff.} & \textbf{P-value} & \textbf{Non-inferior} & \textbf{Margin} & \textbf{N} \\
\hline
PTV HI Mean & 0.004 & 0.00 & Yes & 0.01 & 62 \\
PTV HI Median & 0.004 & 0.00 & Yes & 0.01 & 62 \\
Bladder Mean Dose & 0.392 & 0.00 & Yes & 1.50 & 62 \\
Bladder D50 & 0.346 & 0.00 & Yes & 1.50 & 62 \\
Bladder D2 & 1.135 & 0.03 & Yes & 1.50 & 62 \\
Rectum Mean Dose & 0.031 & 0.00 & Yes & 1.50 & 62 \\
Rectum D50 & -0.126 & 0.00 & Yes & 1.50 & 62 \\
Rectum D2 & 1.137 & 0.03 & Yes & 1.50 & 62 \\
\hline
\end{tabular}
\label{tab:supp_noninferiority}
\end{table}


\supsection{Qualitative Results: Representative Cases}

We depict the DVHs of six representative cases spanning the full spectrum of PTV sizes (0th (minimum), 20th, 40th, 50th (median), 80th, and 100th (maximum) percentiles) of our validation dataset using different dose engines (DL dose, LTBS, Eclipse AAA). 
The results using the Eclipse AcurosXB dose engine are in the main document.

\begin{figure}[H]
\centering
\begin{subfigure}{0.49\textwidth}
    \includegraphics[width=0.95\textwidth, clip, trim=0cm 0cm 5.8cm 0.9cm]{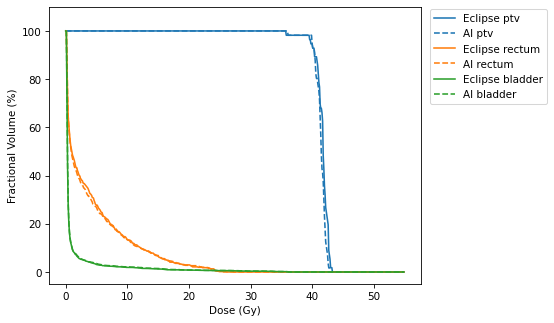}
    \caption{Case 1}
    \label{fig:dvh1}
\end{subfigure}
\hfill
\begin{subfigure}{0.49\textwidth}
    \includegraphics[width=0.95\textwidth, clip, trim=0cm 0cm 5.8cm 0.9cm]{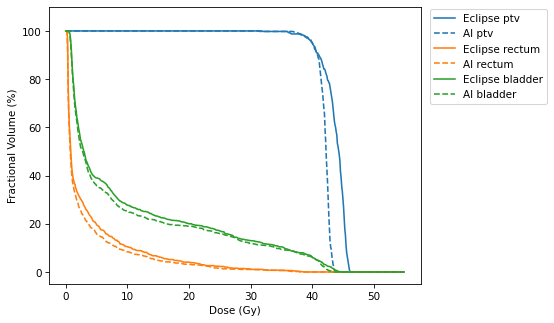}
    \caption{Case 2}
    \label{fig:dvh2}
\end{subfigure}

\begin{subfigure}{0.49\textwidth}
    \includegraphics[width=0.95\textwidth, clip, trim=0cm 0cm 5.8cm 0.9cm]{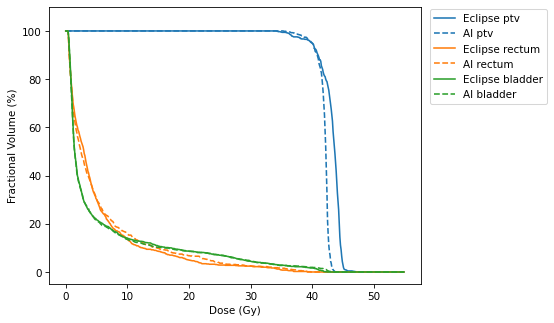}
    \caption{Case 3}
    \label{fig:dvh3}
\end{subfigure}
\hfill
\begin{subfigure}{0.49\textwidth}
    \includegraphics[width=0.95\textwidth, clip, trim=0cm 0cm 5.8cm 0.9cm]{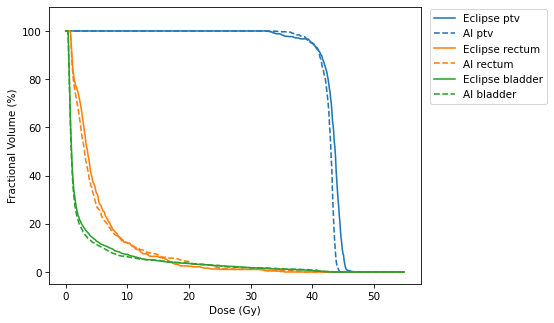}
    \caption{Case 4}
    \label{fig:dvh4}
\end{subfigure}

\begin{subfigure}{0.49\textwidth}
    \includegraphics[width=0.95\textwidth, clip, trim=0cm 0cm 5.8cm 0.9cm]{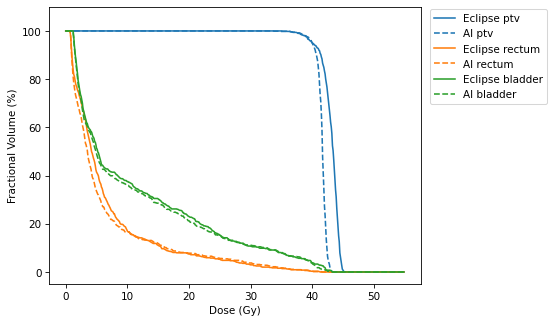}
    \caption{Case 5}
    \label{fig:dvh5}
\end{subfigure}
\hfill
\begin{subfigure}{0.49\textwidth}
    \includegraphics[width=0.95\textwidth, clip, trim=0cm 0cm 5.8cm 0.9cm]{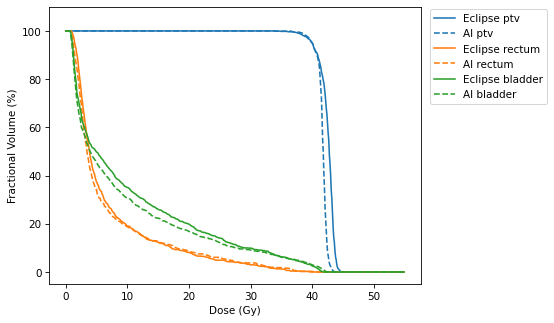}
    \caption{Case 6}
    \label{fig:dvh6}
\end{subfigure}

\vspace{-0.1em}

\centering
\begin{tikzpicture}
\node[
    draw,
    rectangle,
    rounded corners=2pt,
    line width=0.2pt,  
    inner sep=1pt
]{
\begin{tikzpicture}
\matrix[
    column sep=0.5em,  
    row sep=0.2em,  
    nodes={font=\footnotesize, anchor=west}
]{
    \draw[blue, line width=0.8pt] (0,0) -- (0.7,0); &
    \node{PTV (Eclipse)}; &
    \draw[blue, dashed, line width=0.8pt] (0,0) -- (0.7,0); &
    \node{PTV (AI)}; &
    \draw[orange, line width=0.8pt] (0,0) -- (0.7,0); &
    \node{Rectum (Eclipse)};  \\

    \draw[orange, dashed, line width=0.8pt] (0,0) -- (0.7,0); &
    \node{Rectum (AI)}; &
    \draw[green!60!black, line width=0.8pt] (0,0) -- (0.7,0); &
    \node{Bladder (Eclipse)}; &
    \draw[green!60!black, dashed, line width=0.8pt] (0,0) -- (0.7,0); &
    \node{Bladder (AI)}; \\
};
\end{tikzpicture}
};
\end{tikzpicture}

\vspace{0.3em}
\caption{Dose–volume histograms (DVHs) for six cases corresponding to the 0th (minimum), 20th, 40th, 50th (median), 80th, and 100th (maximum) percentiles of PTV size in the validation dataset using DL dose engine.}
\label{fig:dvh_6cases_dl_dose}
\end{figure}

\begin{figure}[H]
\centering
\begin{subfigure}{0.49\textwidth}
    \includegraphics[width=0.95\textwidth, clip, trim=0cm 0cm 5.8cm 0.9cm]{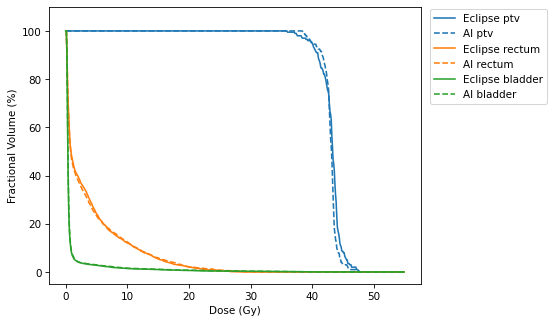} %
    \caption{Case 1}
    \label{fig:dvh1}
\end{subfigure}
\hfill
\begin{subfigure}{0.49\textwidth}
    \includegraphics[width=0.95\textwidth, clip, trim=0cm 0cm 5.8cm 0.9cm]{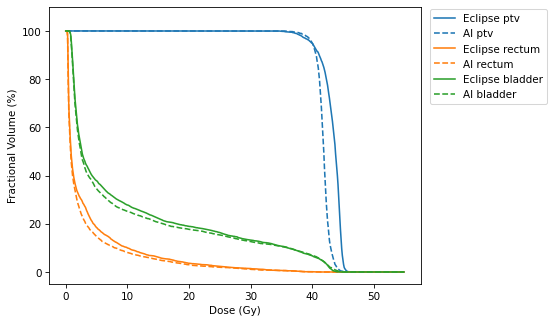}
    \caption{Case 2}
    \label{fig:dvh2}
\end{subfigure}

\begin{subfigure}{0.49\textwidth}
    \includegraphics[width=0.95\textwidth, clip, trim=0cm 0cm 5.8cm 0.9cm]{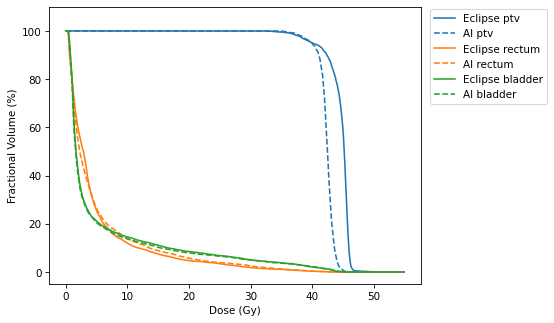}
    \caption{Case 3}
    \label{fig:dvh3}
\end{subfigure}
\hfill
\begin{subfigure}{0.49\textwidth}
    \includegraphics[width=0.95\textwidth, clip, trim=0cm 0cm 5.8cm 0.9cm]{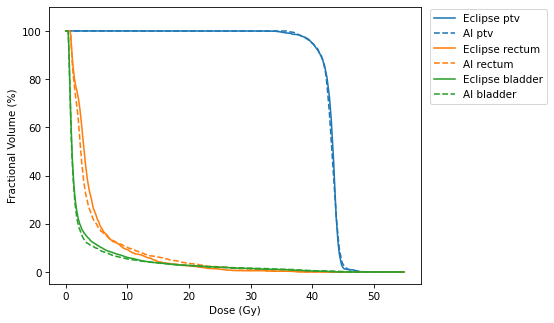}
    \caption{Case 4}
    \label{fig:dvh4}
\end{subfigure}

\begin{subfigure}{0.49\textwidth}
    \includegraphics[width=0.95\textwidth, clip, trim=0cm 0cm 5.8cm 0.9cm]{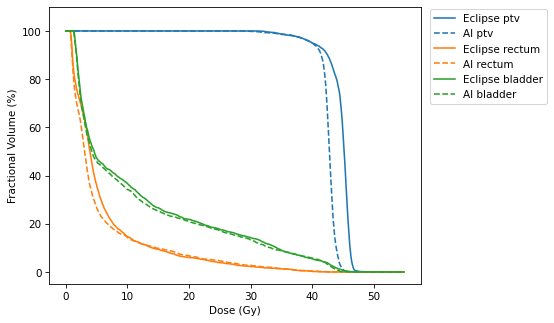}
    \caption{Case 5}
    \label{fig:dvh5}
\end{subfigure}
\hfill
\begin{subfigure}{0.49\textwidth}
    \includegraphics[width=0.95\textwidth, clip, trim=0cm 0cm 5.8cm 0.9cm]{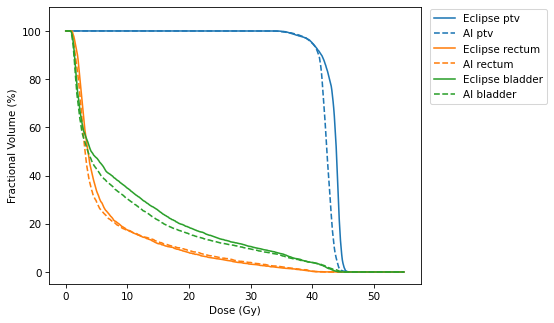}
    \caption{Case 6}
    \label{fig:dvh6}
\end{subfigure}

\vspace{-0.1em}

\centering
\begin{tikzpicture}
\node[
    draw,
    rectangle,
    rounded corners=2pt,
    line width=0.2pt,  
    inner sep=1pt
]{
\begin{tikzpicture}
\matrix[
    column sep=0.5em,  
    row sep=0.2em,  
    nodes={font=\footnotesize, anchor=west}
]{
    \draw[blue, line width=0.8pt] (0,0) -- (0.7,0); &
    \node{PTV (Eclipse)}; &
    \draw[blue, dashed, line width=0.8pt] (0,0) -- (0.7,0); &
    \node{PTV (AI)}; &
    \draw[orange, line width=0.8pt] (0,0) -- (0.7,0); &
    \node{Rectum (Eclipse)};  \\

    \draw[orange, dashed, line width=0.8pt] (0,0) -- (0.7,0); &
    \node{Rectum (AI)}; &
    \draw[green!60!black, line width=0.8pt] (0,0) -- (0.7,0); &
    \node{Bladder (Eclipse)}; &
    \draw[green!60!black, dashed, line width=0.8pt] (0,0) -- (0.7,0); &
    \node{Bladder (AI)}; \\
};
\end{tikzpicture}
};
\end{tikzpicture}

\vspace{0.3em}
\caption{Dose–volume histograms (DVHs) for six cases corresponding to the 0th (minimum), 20th, 40th, 50th (median), 80th, and 100th (maximum) percentiles of PTV size in the validation dataset using LTBS dose engine.}
\label{fig:dvh_6cases_ltbs}
\end{figure}

\begin{figure}[H]
\centering
\begin{subfigure}{0.49\textwidth}
    \includegraphics[width=0.95\textwidth, clip, trim=0cm 0cm 5.8cm 0.9cm]{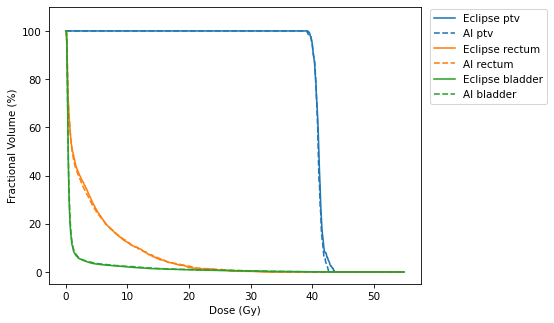} %
    \caption{Case 1}
    \label{fig:dvh1}
\end{subfigure}
\hfill
\begin{subfigure}{0.49\textwidth}
    \includegraphics[width=0.95\textwidth, clip, trim=0cm 0cm 5.8cm 0.9cm]{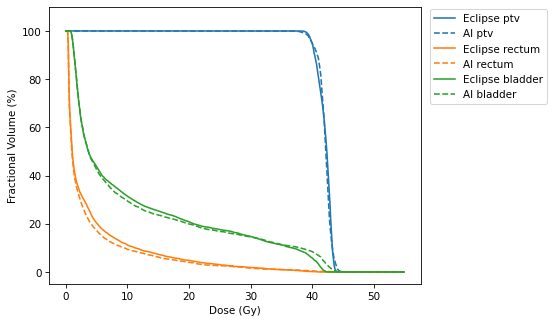}
    \caption{Case 2}
    \label{fig:dvh2}
\end{subfigure}

\begin{subfigure}{0.49\textwidth}
    \includegraphics[width=0.95\textwidth, clip, trim=0cm 0cm 5.8cm 0.9cm]{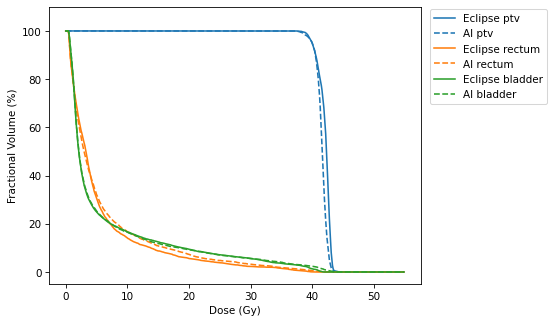}
    \caption{Case 3}
    \label{fig:dvh3}
\end{subfigure}
\hfill
\begin{subfigure}{0.49\textwidth}
    \includegraphics[width=0.95\textwidth, clip, trim=0cm 0cm 5.8cm 0.9cm]{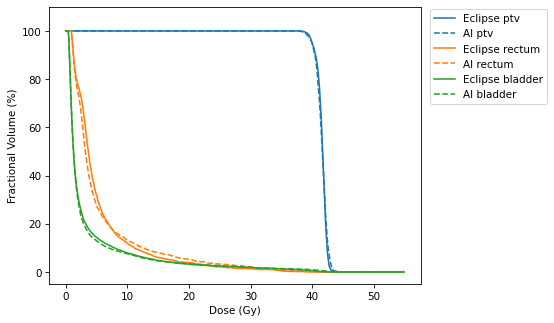}
    \caption{Case 4}
    \label{fig:dvh4}
\end{subfigure}

\begin{subfigure}{0.49\textwidth}
    \includegraphics[width=0.95\textwidth, clip, trim=0cm 0cm 5.8cm 0.9cm]{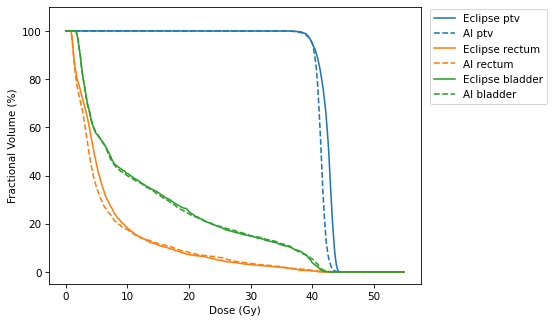}
    \caption{Case 5}
    \label{fig:dvh5}
\end{subfigure}
\hfill
\begin{subfigure}{0.49\textwidth}
    \includegraphics[width=0.95\textwidth, clip, trim=0cm 0cm 5.8cm 0.9cm]{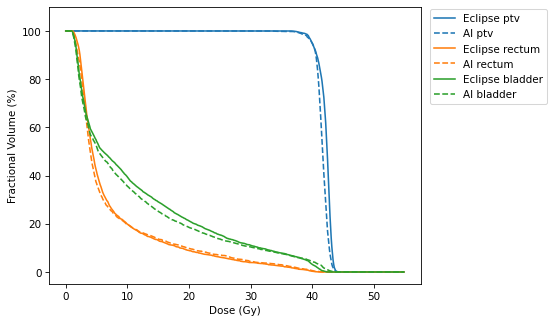}
    \caption{Case 6}
    \label{fig:dvh6}
\end{subfigure}

\vspace{-0.1em}

\centering
\begin{tikzpicture}
\node[
    draw,
    rectangle,
    rounded corners=2pt,
    line width=0.2pt,  
    inner sep=1pt
]{
\begin{tikzpicture}
\matrix[
    column sep=0.5em,  
    row sep=0.2em,  
    nodes={font=\footnotesize, anchor=west}
]{
    \draw[blue, line width=0.8pt] (0,0) -- (0.7,0); &
    \node{PTV (Eclipse)}; &
    \draw[blue, dashed, line width=0.8pt] (0,0) -- (0.7,0); &
    \node{PTV (AI)}; &
    \draw[orange, line width=0.8pt] (0,0) -- (0.7,0); &
    \node{Rectum (Eclipse)};  \\

    \draw[orange, dashed, line width=0.8pt] (0,0) -- (0.7,0); &
    \node{Rectum (AI)}; &
    \draw[green!60!black, line width=0.8pt] (0,0) -- (0.7,0); &
    \node{Bladder (Eclipse)}; &
    \draw[green!60!black, dashed, line width=0.8pt] (0,0) -- (0.7,0); &
    \node{Bladder (AI)}; \\
};
\end{tikzpicture}
};
\end{tikzpicture}

\vspace{0.3em}
\caption{Dose–volume histograms (DVHs) for six cases corresponding to the 0th (minimum), 20th, 40th, 50th (median), 80th, and 100th (maximum) percentiles of PTV size in the validation dataset using Eclipse AAA dose engine.}
\label{fig:dvh_6cases_AAA}
\end{figure}

We depict the mean curves accross patients for our \AIRT method and the Eclipse optimized RapidPlan evaluated on the DL dose engine in Figure~\ref{fig:dvh_mean2cases_dl_dose}, evaluated on LTBS dose engine in Figure~\ref{fig:dvh_mean2cases_ltbs}, 
evaluated on Eclipse AAA in Figure~\ref{fig:dvh_mean2cases_aaa} and evaluated on Eclipse AcurosXB in Figure~\ref{fig:dvh_mean2cases_acuros}.

Figure~\ref{fig:dose_cases4_6_DL} shows the dose distribution of the median validation case (by PTV volume) using DL dose evaluation, complementing the DVH comparisons in Figure~\ref{fig:dvh_6cases_dl_dose}.

Figure~\ref{fig:dose_cases4_6_LTBS} shows the dose distribution of the median validation case (by PTV volume) using LTBS dose evaluation, complementing the DVH comparisons in Figure~\ref{fig:dvh_6cases_ltbs}.

Figure~\ref{fig:dose_cases4_6_AAA} shows the dose distribution of the median validation case (by PTV volume) using Eclipse AAA dose evaluation, complementing the DVH comparisons in Figure~\ref{fig:dvh_6cases_AAA}.

Figure~\ref{fig:dose_cases4_6_Acuros} shows the dose distribution of the median validation case (by PTV volume) using Eclipse AcurosXB evaluation, complementing the DVH comparisons in Figure~\ref{fig:dvh_6cases_Acuros} of the main manuscript.

\begin{figure}[H]
\centering
\begin{subfigure}{0.49\textwidth}
    \includegraphics[width=\textwidth]{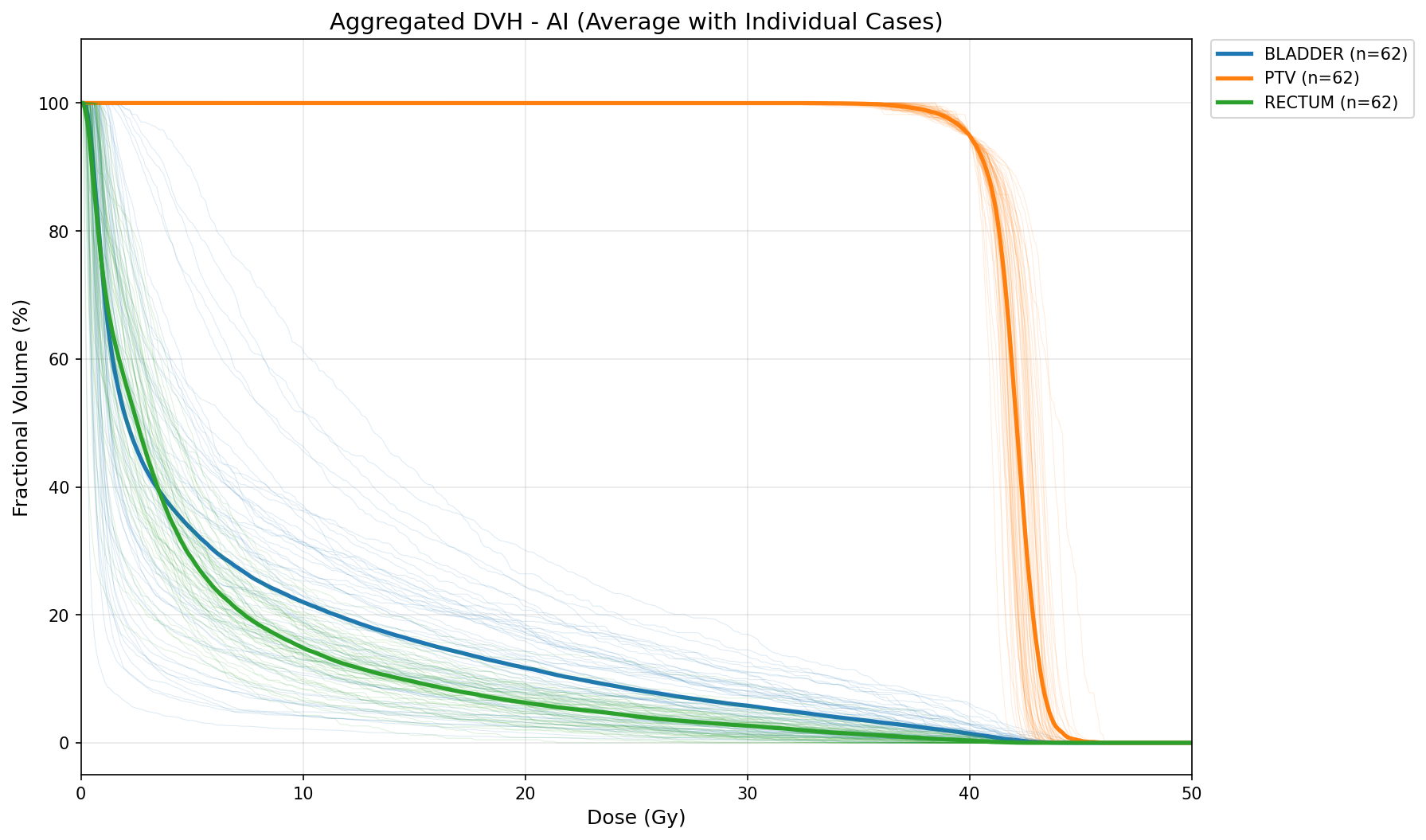}
    \caption{DVHs curves of our \AIRT method.}
    \label{fig:dvh_mean1}
\end{subfigure}
\hfill
\begin{subfigure}{0.49\textwidth}
    \includegraphics[width=\textwidth]{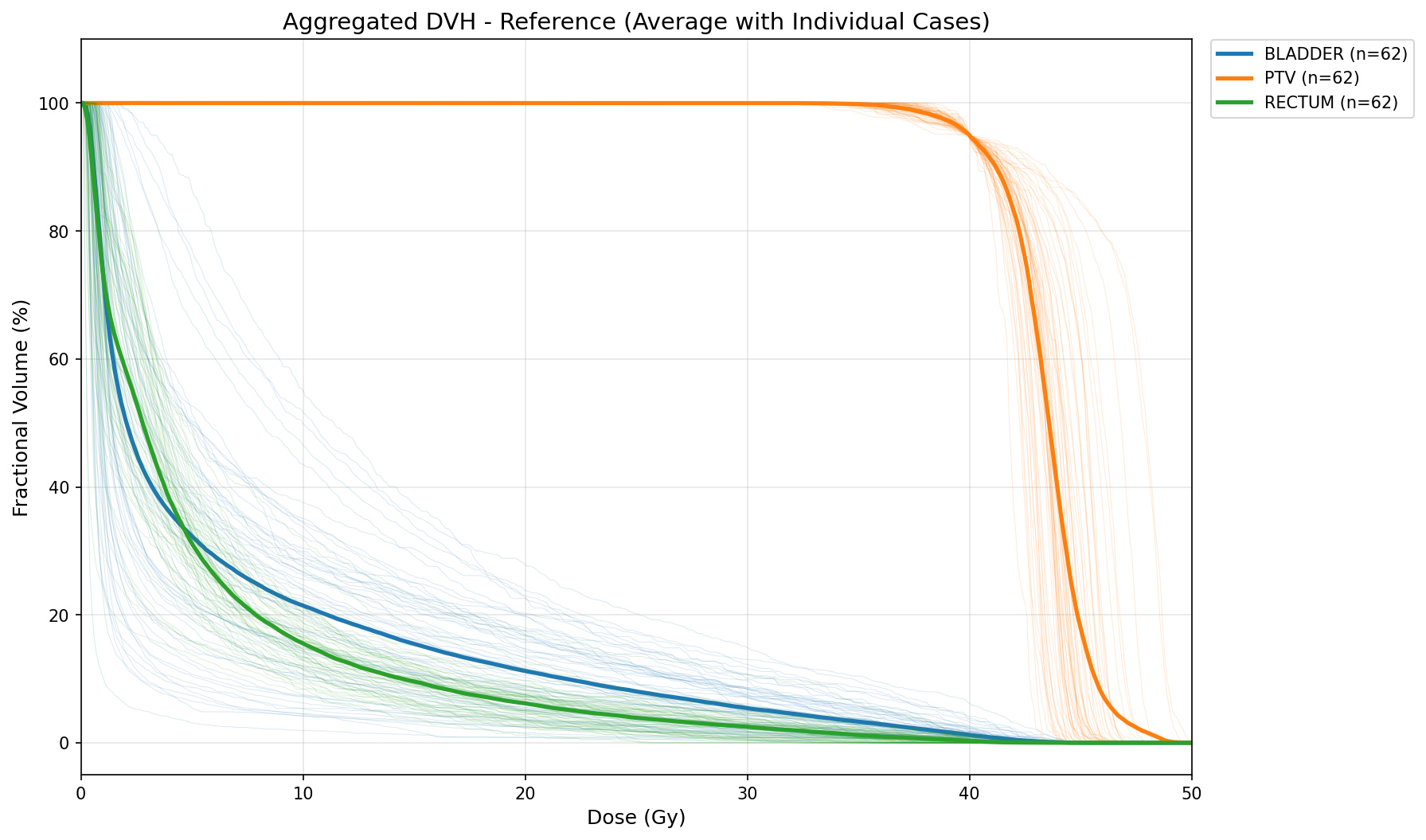}
    \caption{DVHs curves of Eclipse optimized plans.}
    \label{fig:dvh_mean2}
\end{subfigure}

\caption{DVHs curves accross all the validation patients showing dose distributions for PTV and OARs using DL dose engine. In bold are the mean curves accross patients.}
\label{fig:dvh_mean2cases_dl_dose}
\end{figure}

\begin{figure}[H]
\centering
\begin{subfigure}{0.49\textwidth}
    \includegraphics[width=\textwidth]{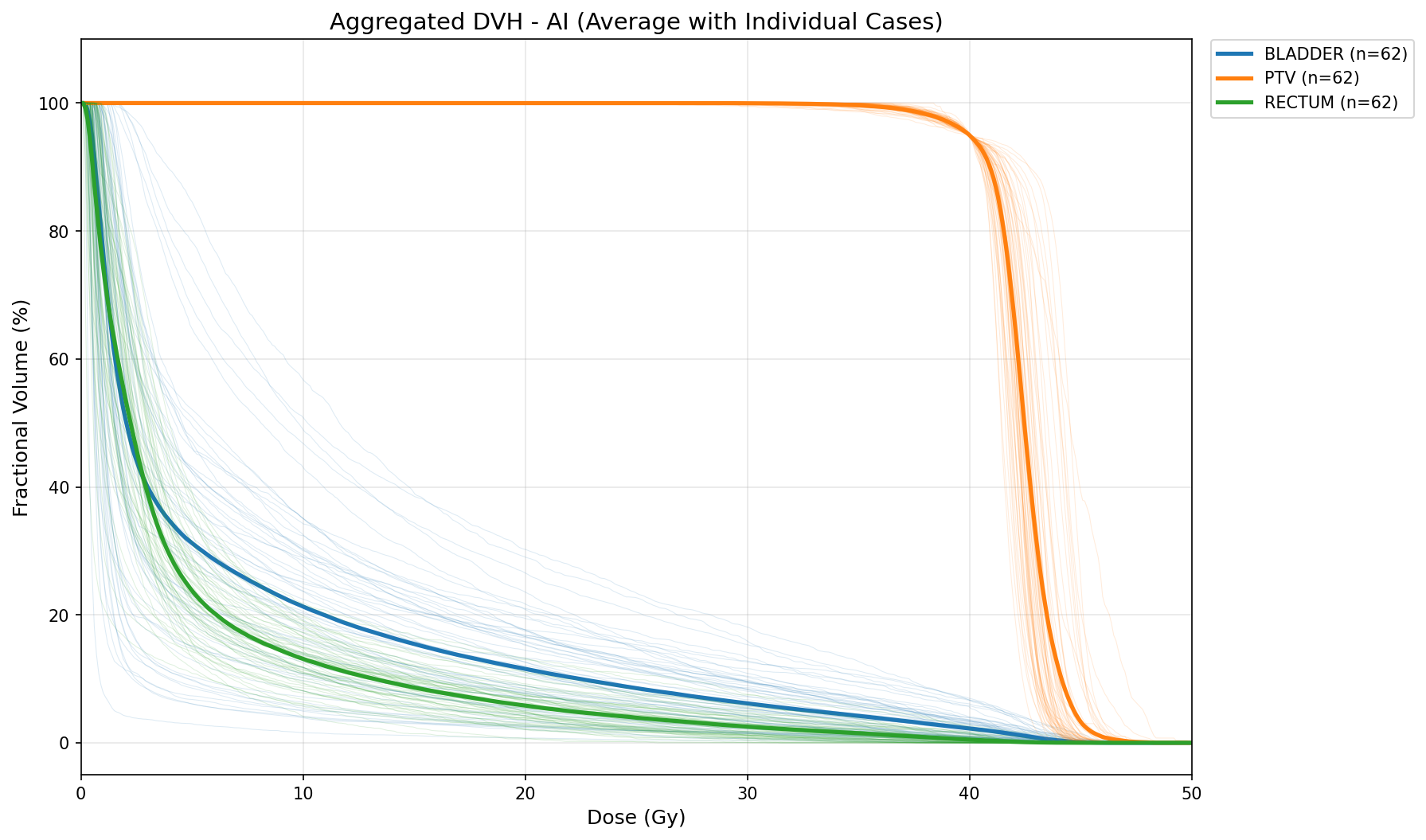}
    \caption{DVHs curves of our \AIRT method.}
    \label{fig:dvh_mean1}
\end{subfigure}
\hfill
\begin{subfigure}{0.49\textwidth}
    \includegraphics[width=\textwidth]{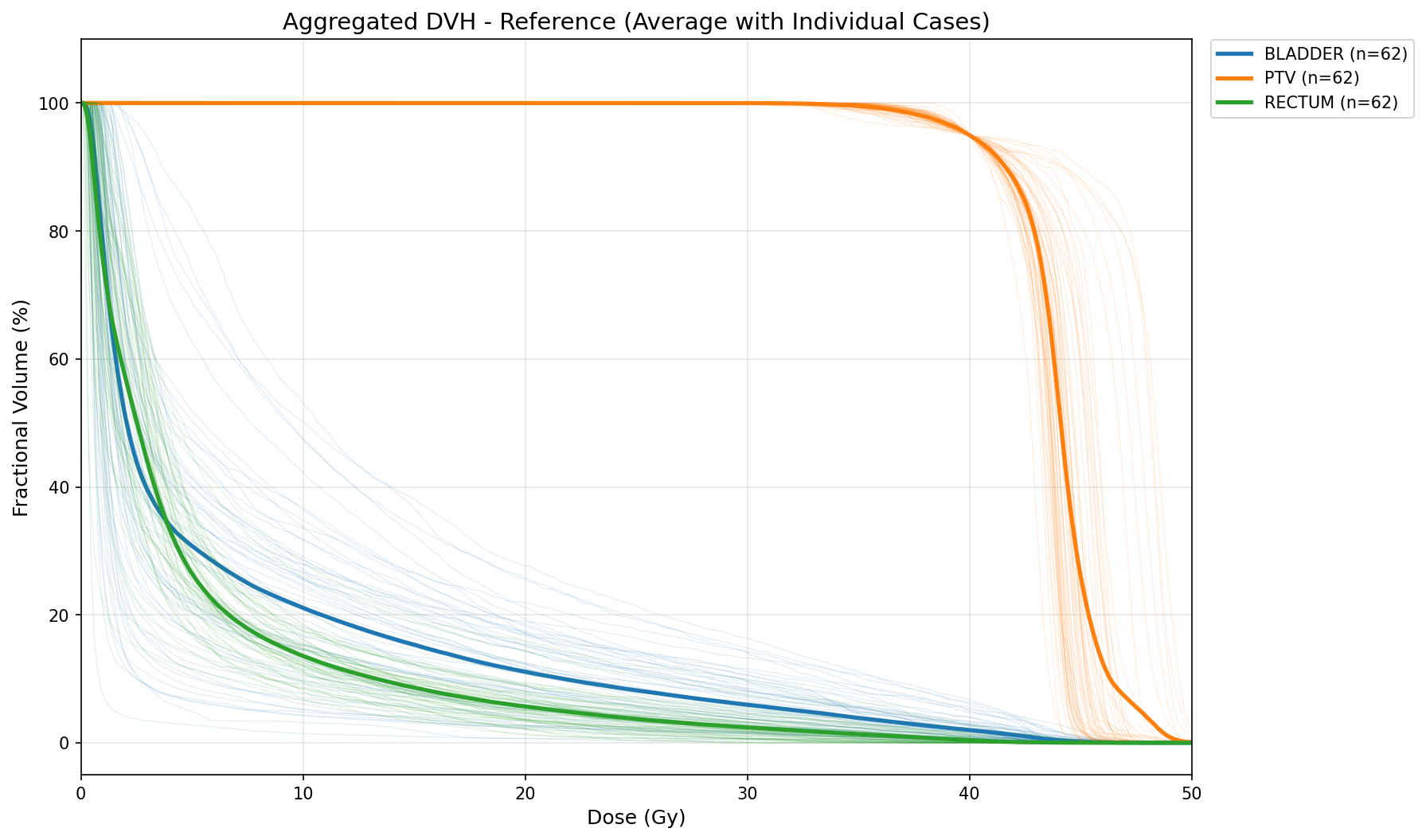}
    \caption{DVHs curves of Eclipse optimized plans.}
    \label{fig:dvh_mean2}
\end{subfigure}

\caption{DVHs curves accross all the validation patients showing dose distributions for PTV and OARs using LTBS dose engine. In bold are the mean curves accross patients.}
\label{fig:dvh_mean2cases_ltbs}
\end{figure}

\begin{figure}[H]
\centering
\begin{subfigure}{0.49\textwidth}
    \includegraphics[width=\textwidth]{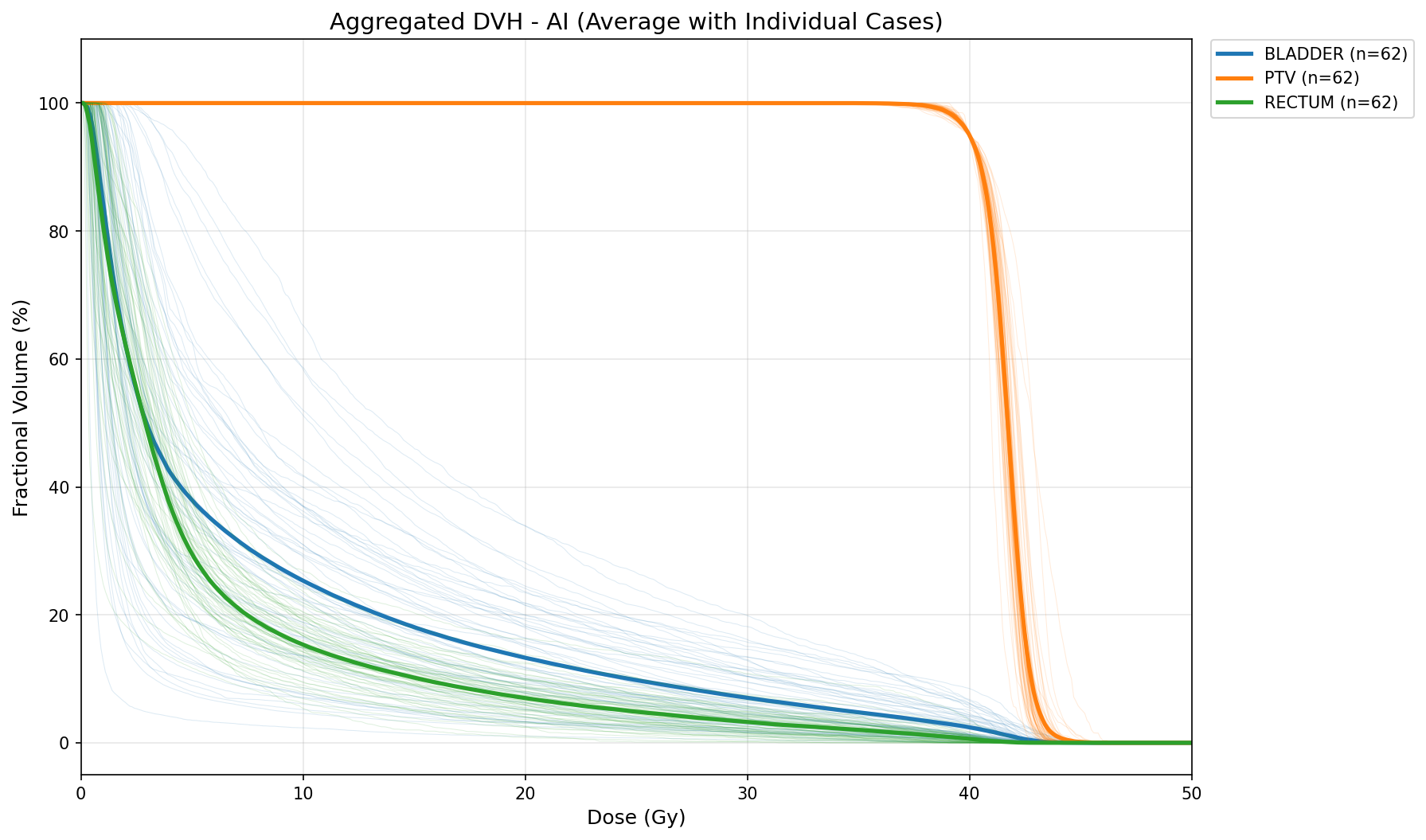}
    \caption{DVHs curves of our \AIRT method.}
    \label{fig:dvh_mean1}
\end{subfigure}
\hfill
\begin{subfigure}{0.49\textwidth}
    \includegraphics[width=\textwidth]{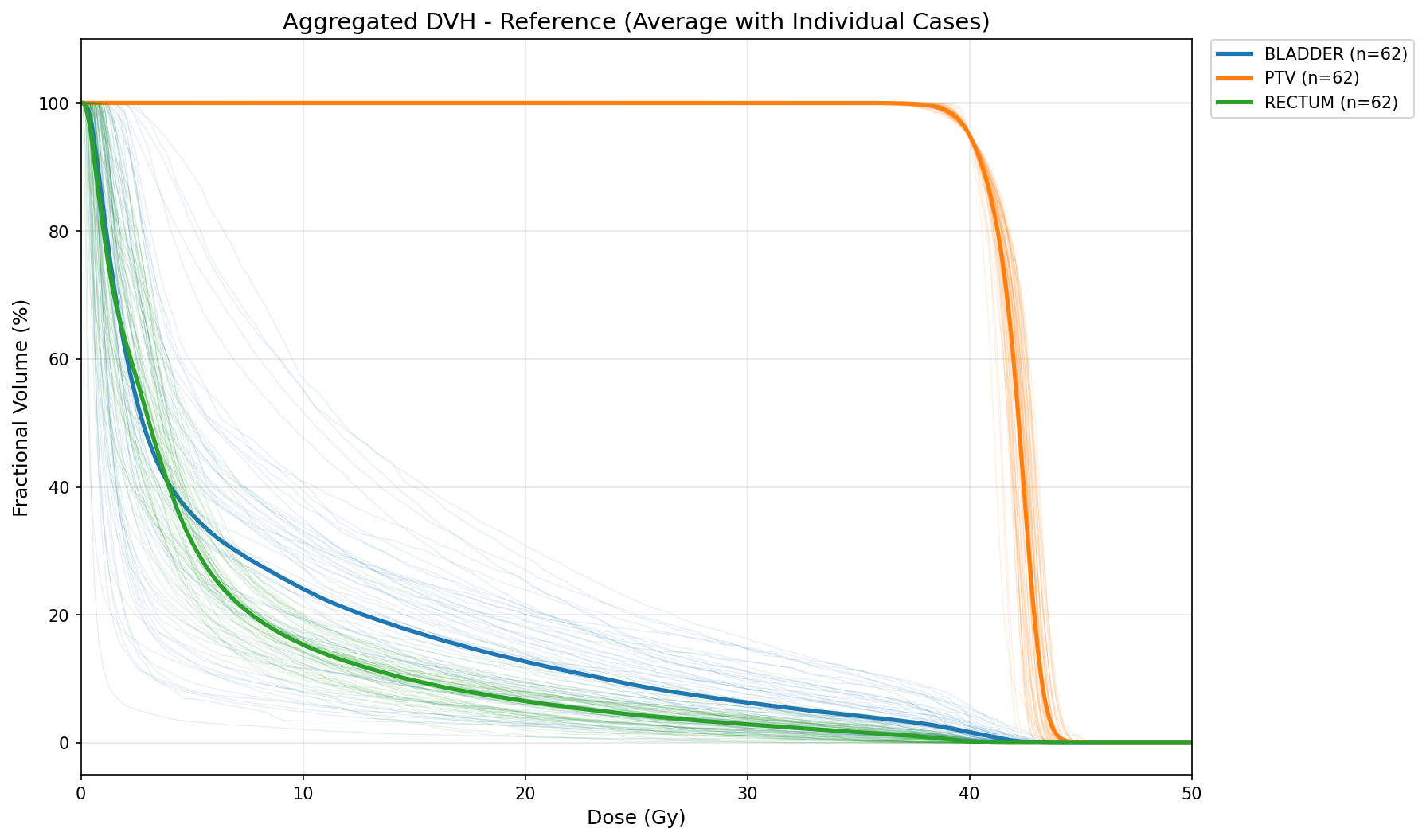}
    \caption{DVHs curves of Eclipse optimized plans.}
    \label{fig:dvh_mean2}
\end{subfigure}

\caption{DVHs curves accross all the validation patients showing dose distributions for PTV and OARs using Eclipse AAA dose engine. In bold are the mean curves accross patients.}
\label{fig:dvh_mean2cases_aaa}
\end{figure}

\begin{figure}[H]
\centering
\begin{subfigure}{0.49\textwidth}
    \includegraphics[width=\textwidth]{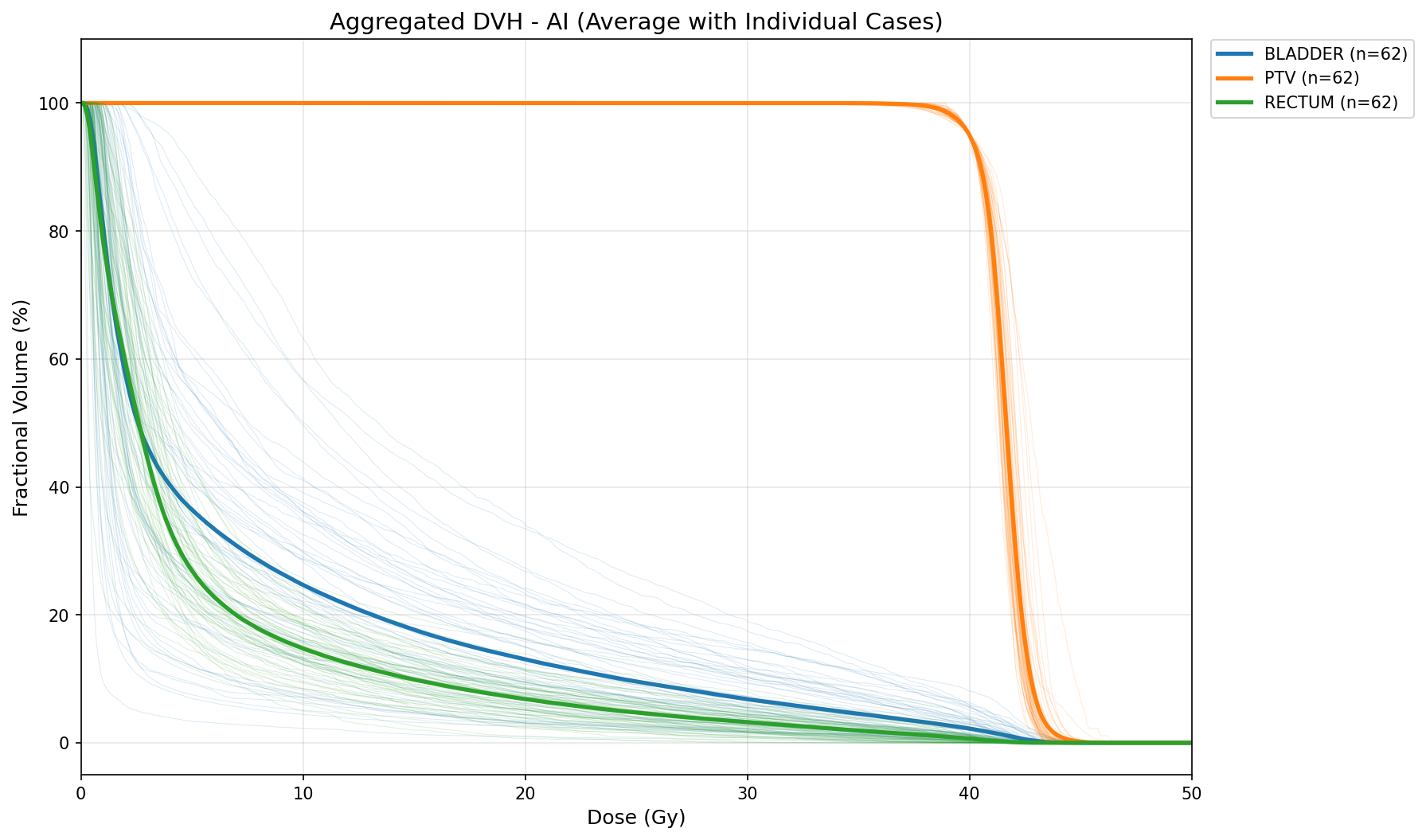}
    \caption{DVHs curves of our \AIRT method.}
    \label{fig:dvh_mean1}
\end{subfigure}
\hfill
\begin{subfigure}{0.49\textwidth}
    \includegraphics[width=\textwidth]{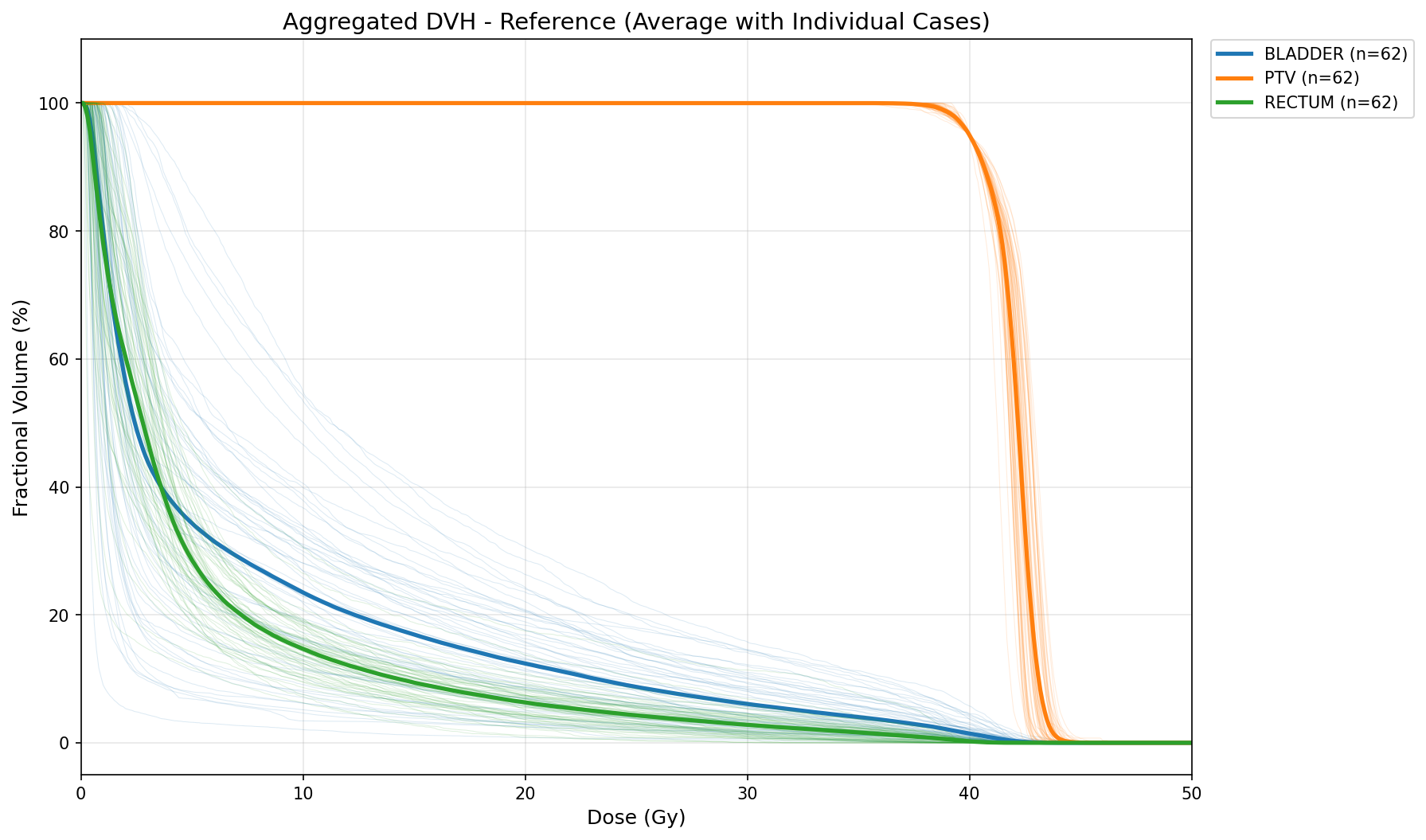}
    \caption{DVHs curves of Eclipse optimized plans.}
    \label{fig:dvh_mean2}
\end{subfigure}

\caption{DVHs curves accross all the validation patients showing dose distributions for PTV and OARs using Eclipse Acuros dose engine. In bold are the mean curves accross patients.}
\label{fig:dvh_mean2cases_acuros}
\end{figure}

\begin{figure}[H]
\centering

\begin{subfigure}{\textwidth}
    \includegraphics[width=0.49\textwidth]{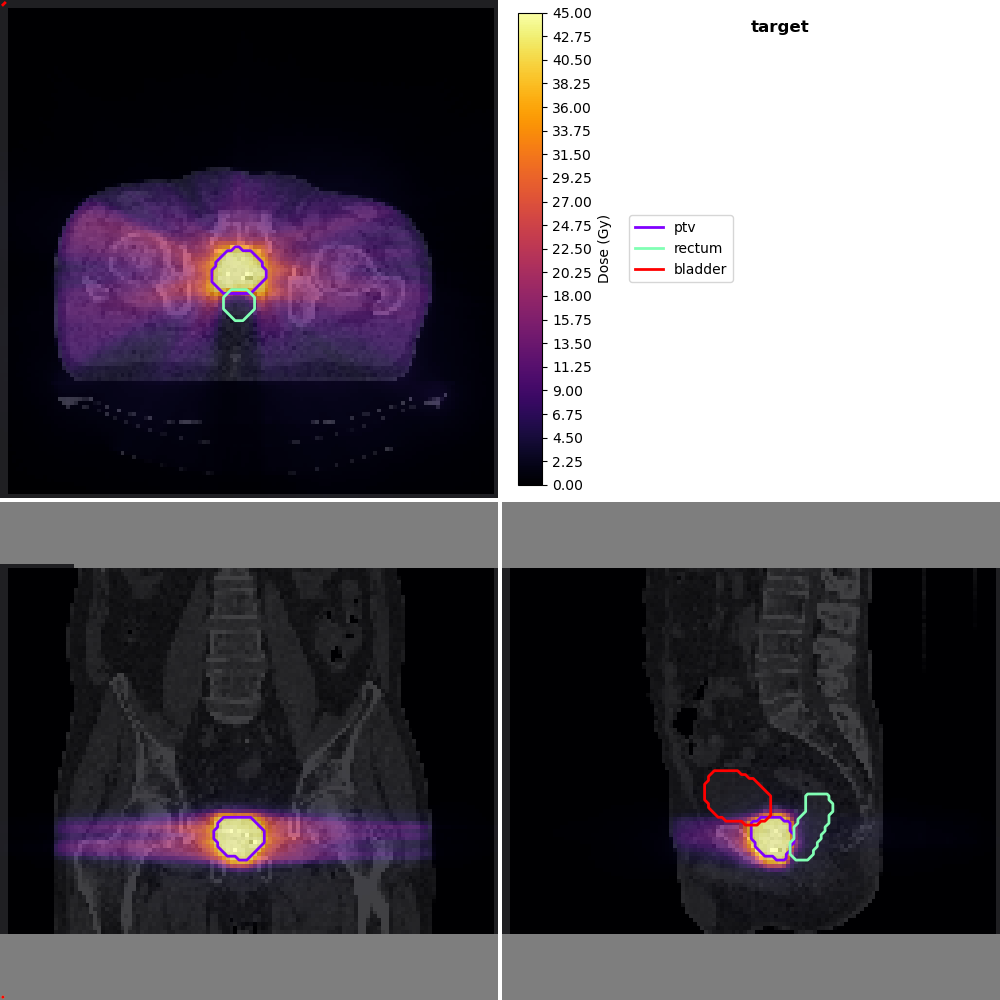}
    \includegraphics[width=0.49\textwidth]{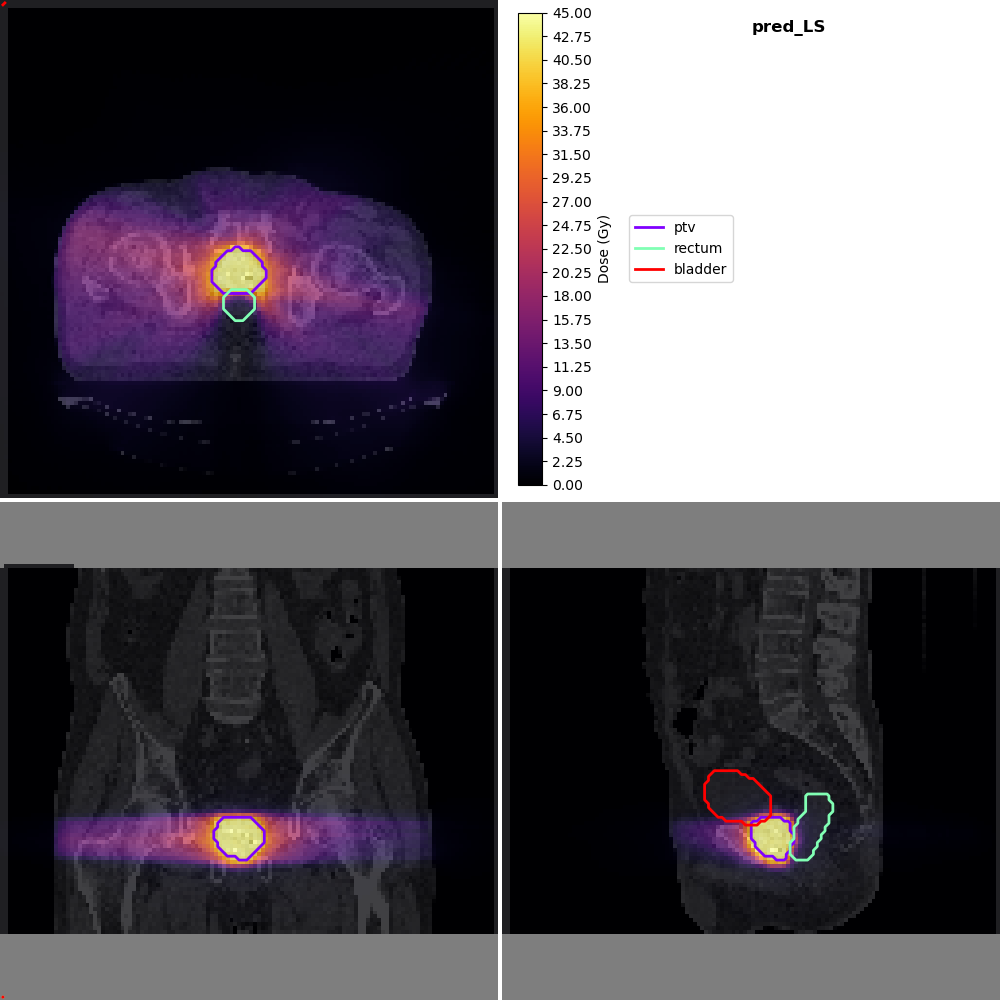}
    \caption{Case 4: Left = Target dose, Right = \AIRT prediction}
    \label{fig:dose_case4}
\end{subfigure}

%
%
%

\caption{Dose distributions evaluated with the DL dose engine for the median validation case (Case 4, as shown in the corresponding DVH figure). Left = reference dose (Eclipse/target), Right = \AIRT predicted dose. CT slices and contours overlaid.}
\label{fig:dose_cases4_6_DL}
\end{figure}

\begin{figure}[H]
\centering

\begin{subfigure}{\textwidth}
    \includegraphics[width=0.49\textwidth]{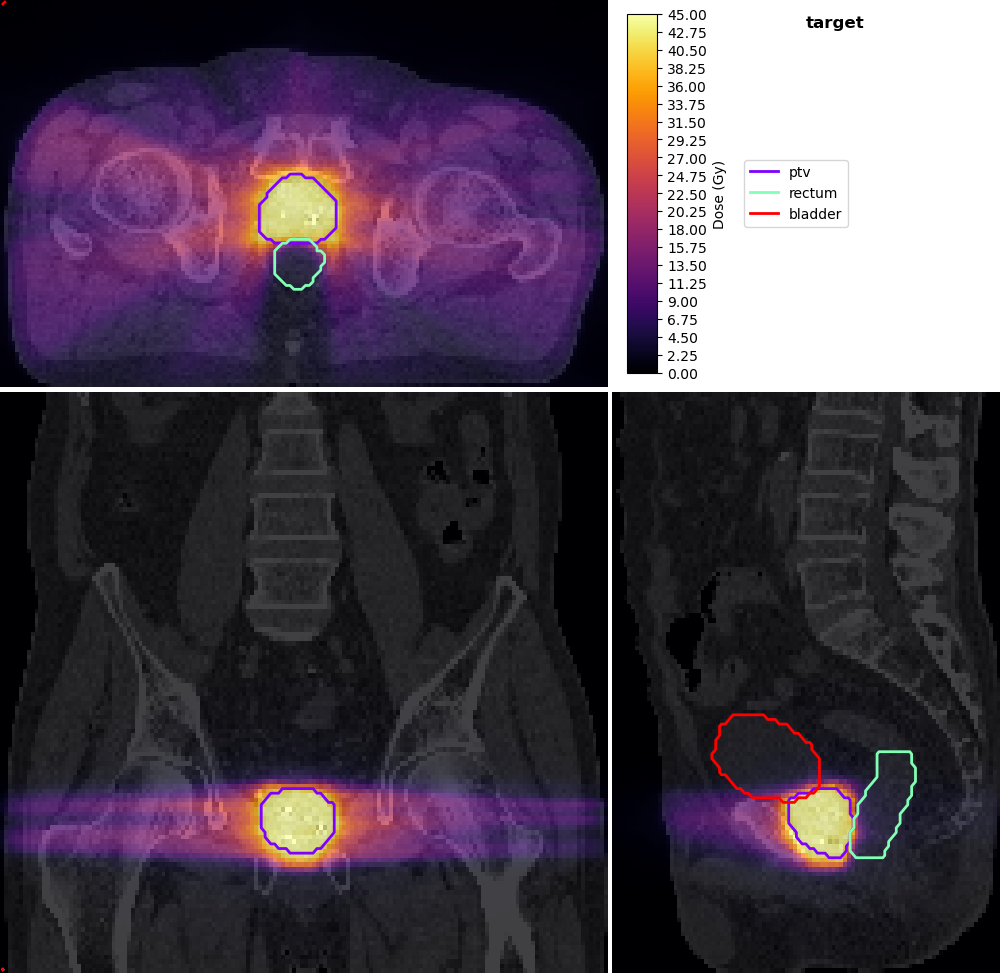}
    \includegraphics[width=0.49\textwidth]{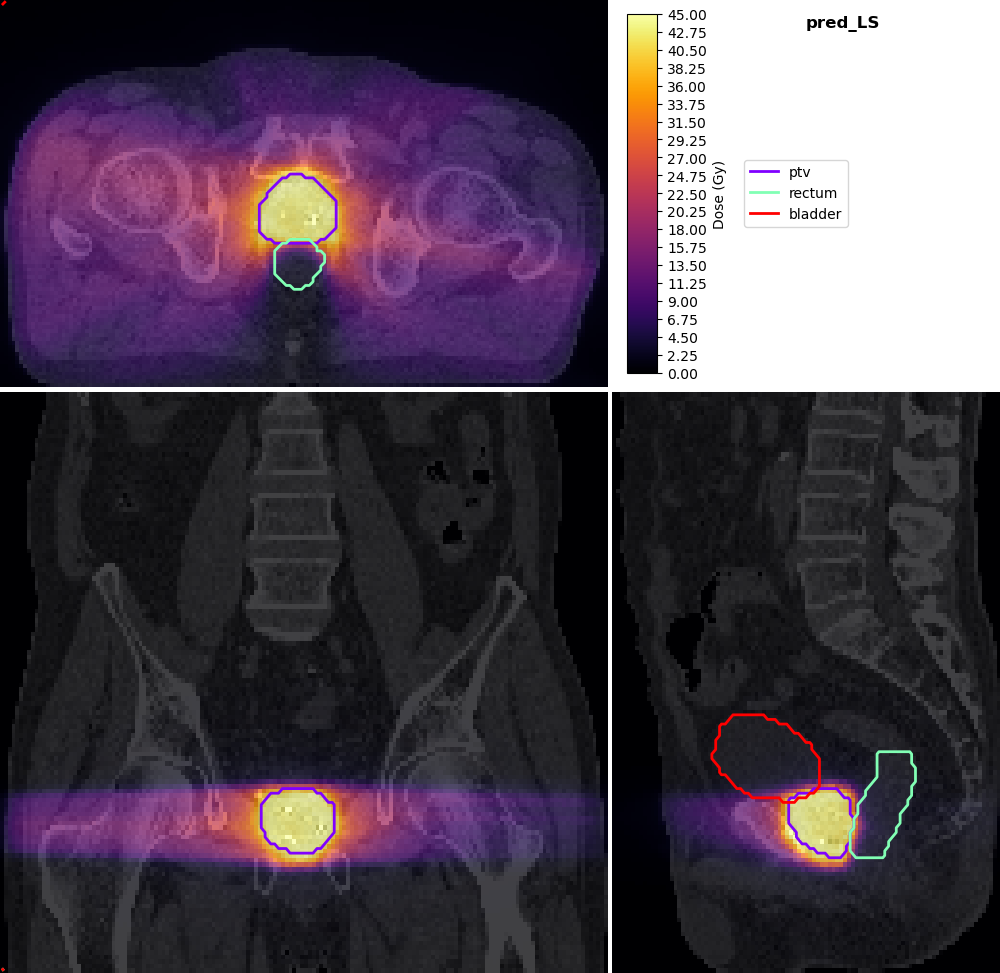}
    \caption{Case 4: Left = Target dose, Right = \AIRT prediction}
    \label{fig:dose_case4}
\end{subfigure}

%
%
%

\caption{Dose distributions evaluated with the LTBS dose engine for the median validation case (Case 4, as shown in the corresponding DVH figure). Left = reference dose (Eclipse/target), Right = \AIRT predicted dose. CT slices and contours overlaid.}
\label{fig:dose_cases4_6_LTBS}
\end{figure}

\begin{figure}[H]
\centering

\begin{subfigure}{\textwidth}
    \includegraphics[width=0.49\textwidth]{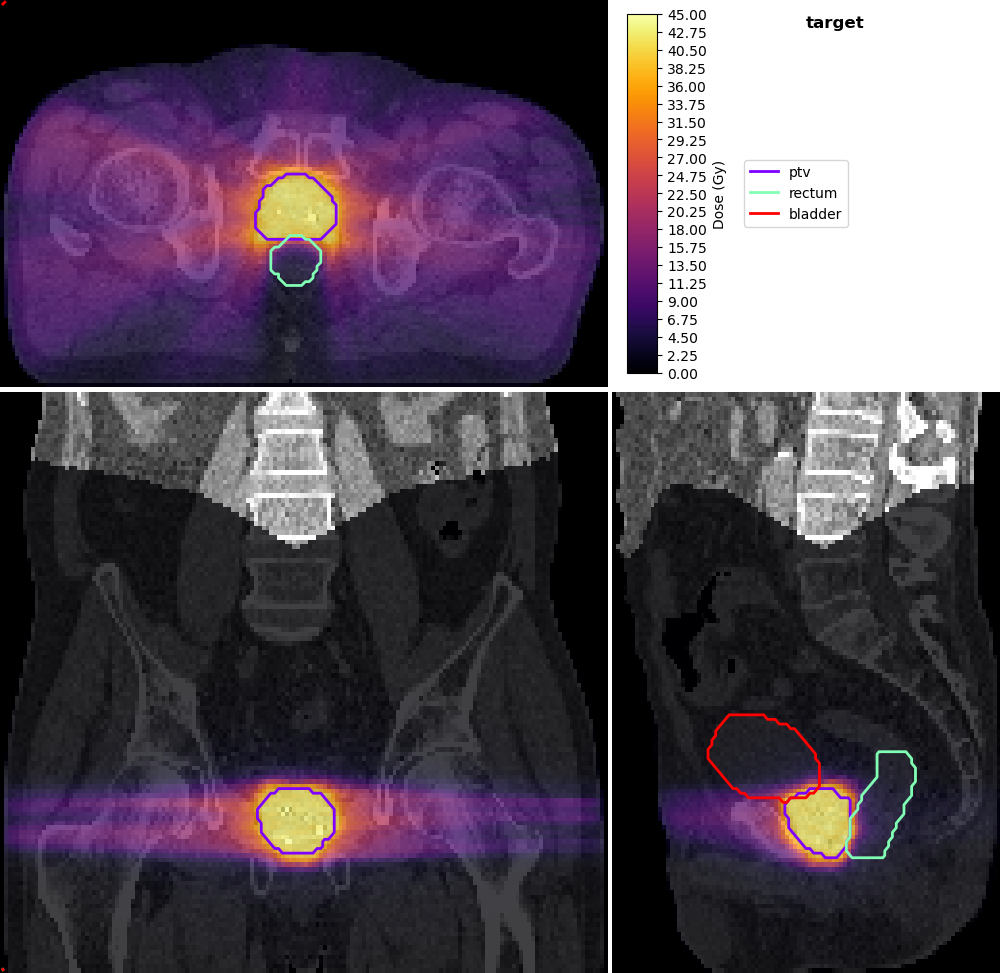}
    \includegraphics[width=0.49\textwidth]{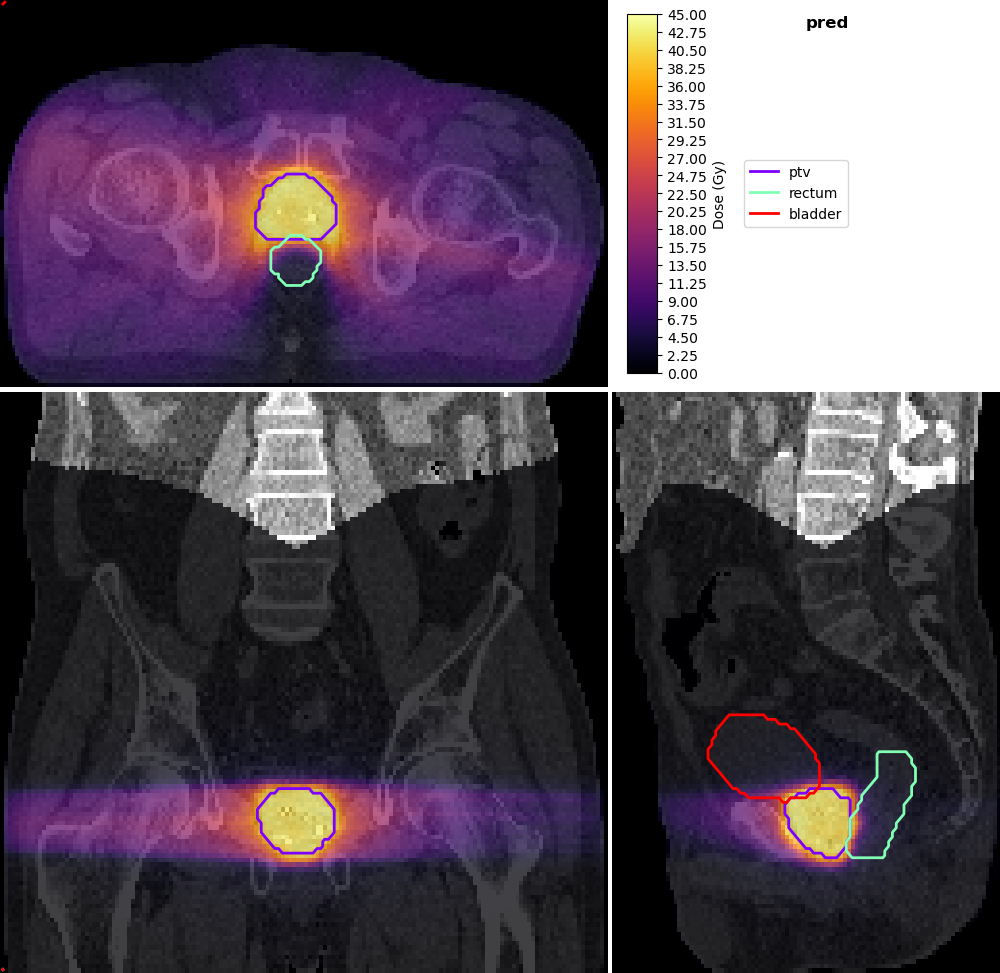}
    \caption{Case 4: Left = Target dose, Right = \AIRT prediction}
    \label{fig:dose_case4}
\end{subfigure}

%
%
%

\caption{Dose distributions evaluated with the Eclipse AAA dose engine for the median validation case (Case 4, as shown in the corresponding DVH figure). Left = reference dose (Eclipse/target), Right = \AIRT predicted dose. CT slices and contours overlaid.}
\label{fig:dose_cases4_6_AAA}
\end{figure}

\begin{figure}[H]
\centering

\begin{subfigure}{\textwidth}
    \includegraphics[width=0.49\textwidth]{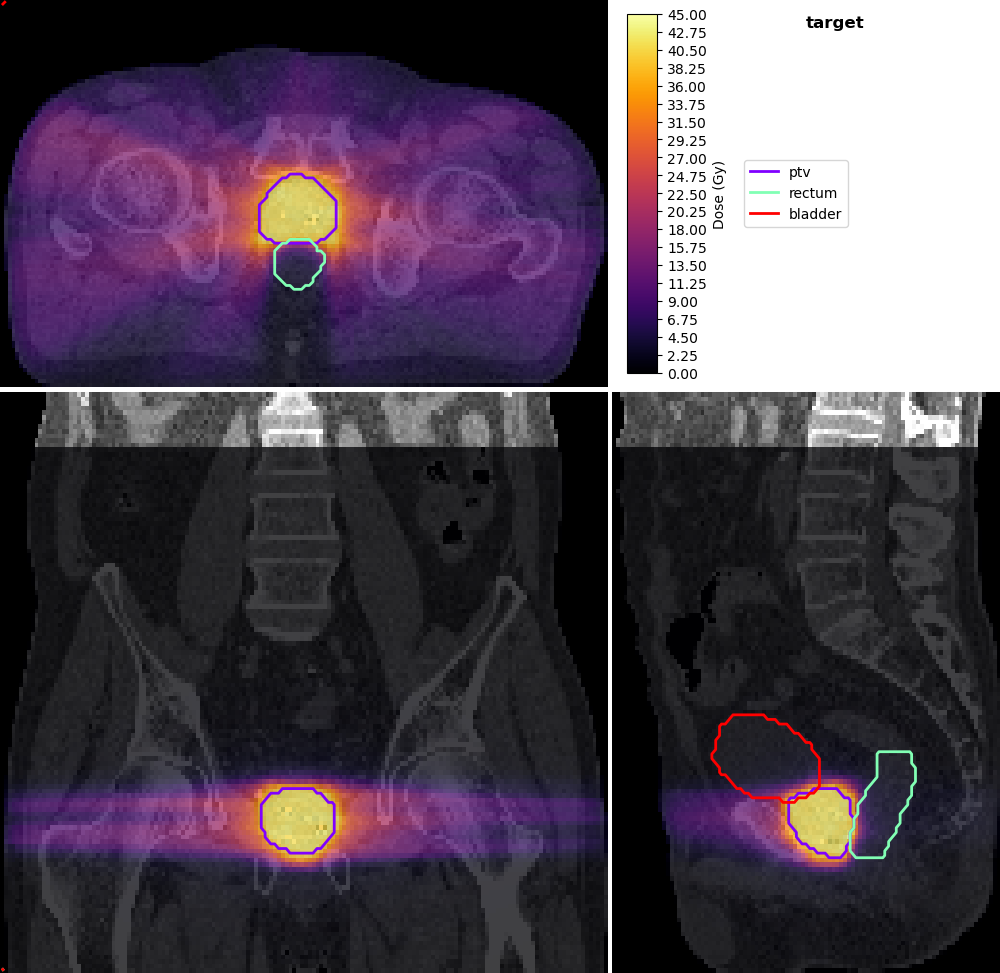}
    \includegraphics[width=0.49\textwidth]{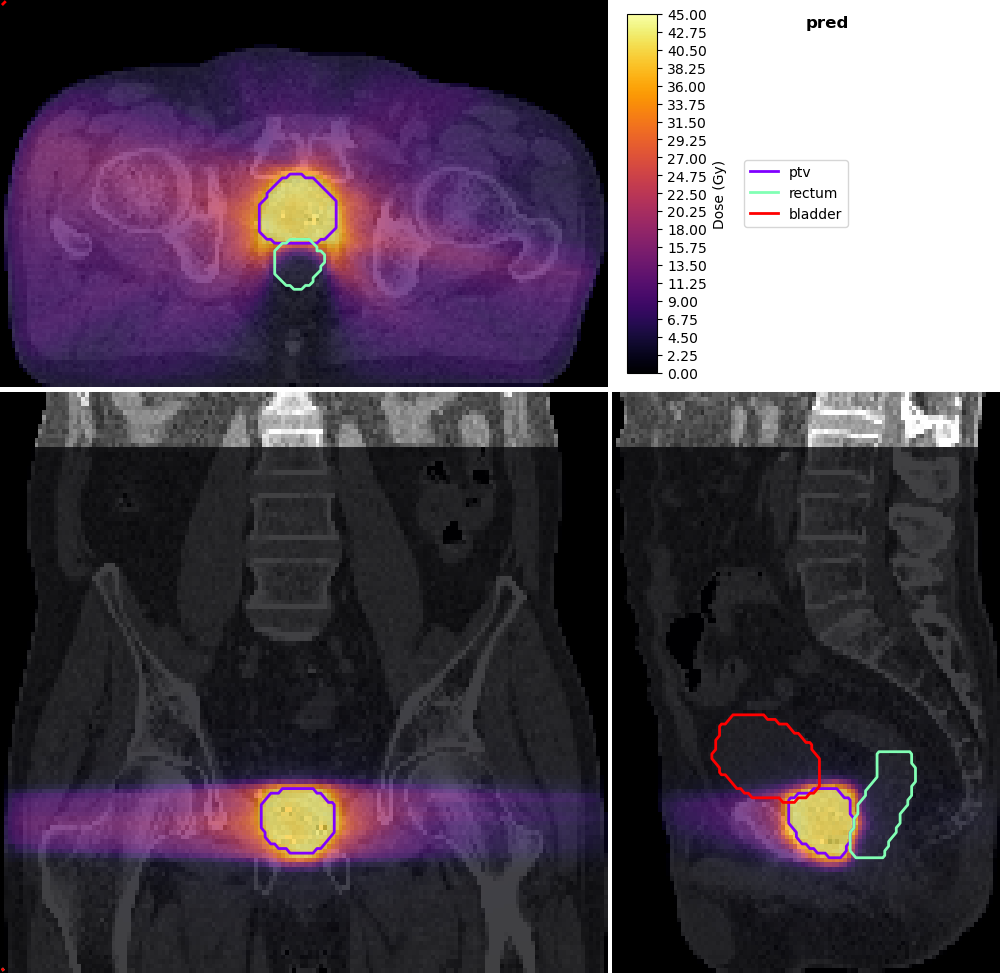}
    \caption{Case 4: Left = Target dose, Right = \AIRT prediction}
    \label{fig:dose_case4}
\end{subfigure}

%
%
%

\caption{Dose distributions evaluated with the Eclipse AcurosXB dose engine for the median validation case (Case 4, as shown in the corresponding DVH figure). Left = reference dose (Eclipse/target), Right = \AIRT predicted dose. CT slices and contours overlaid.}
\label{fig:dose_cases4_6_Acuros}
\end{figure}

\subsection*{Supplementary DVHs demonstrating user-controlled OAR sparing}


To illustrate the adaptability of our end-to-end AI method to user-controlled inputs, we show the average DVH curves across the entire validation dataset for the PTV, rectum, and bladder for various user inputs $s_b$ (for bladder) and $s_r$ (for rectum), controlling the degree of additional dose reduction in each OAR. 
Fig.~\ref{fig:oar_adapt_dvh_supp} illustrates the achievable dose tradeoffs, evaluated using the Dl dose engine, with more aggressive attenuation as in the main document.

\begin{figure}[H]
    \centering
    \includegraphics[width=.99\linewidth]{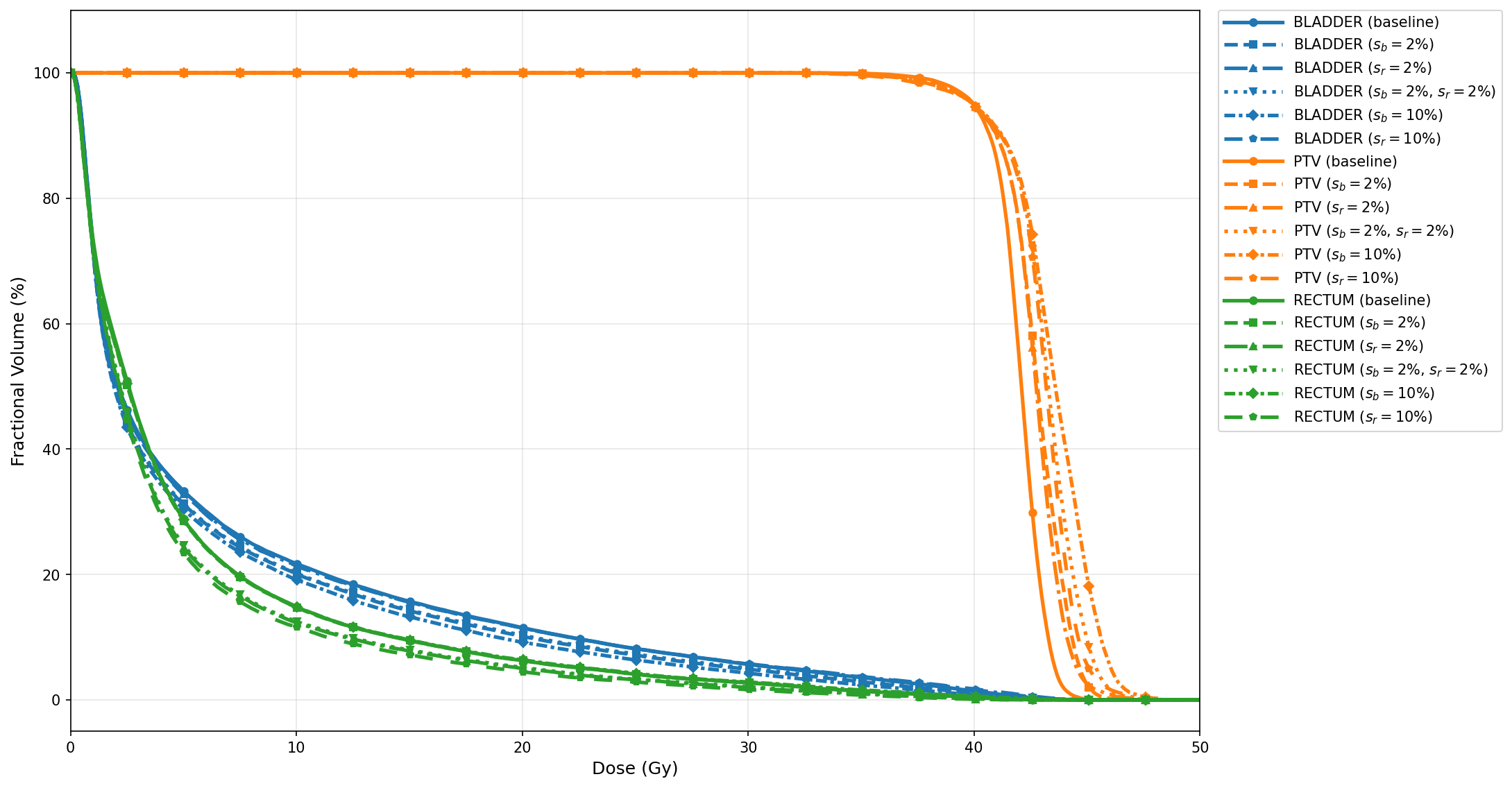}
    \caption{
Mean DVHs for the PTV, bladder, and rectum, averaged over the entire validation set.
Suppression factors are reported as a percentage of the predicted baseline dose. 
The “baseline” curve corresponds to no additional OAR control ($s_b = s_r = 0$).
The figure shows that parameters $s_b$ and $s_r$ independently reduce the OAR dose, with an expected tradeoff on the PTV homogeneity, which worsens as OAR sparing increases.
}
    \label{fig:oar_adapt_dvh_supp}
\end{figure}

\supsection{Computational Performance}

Table \ref{tab:timing} reports the module-wise computational time of the proposed end-to-end VMAT planning pipeline (organ auto-contouring and DICOM export are not included). 
All runtime measurements were performed on a single NVIDIA A100 GPU (80 GB). 
Reported values are the average steady-state inference runtime per patient over the \Nval validation cases.
Two dummy cases were run before timing to account for the one-time initialization overheads associated with model loading, CUDA context creation and GPU kernel autotuning and caching.
The most computationally intensive step is deep-learning-based dose calculation (654~ms). 
As a comparison, the corresponding optimization in Eclipse takes several (3--5) minutes per plan.




\begin{table}[htbp]
\centering
\caption{Module-wise compute time breakdown for the proposed pipeline.}
\label{tab:timing}
\begin{tabularx}{\textwidth}{l>{\raggedright\arraybackslash}Xc}
\toprule
\textbf{Module} & \textbf{Notes} & \textbf{Compute Time (ms)} \\
\midrule
DoseProposer pre-processing & Prepare/crop CT and structure tensors & 35 \\
DoseProposer & Initial dose prediction & 4 \\
BEV Projection (1st pass) & Beam's eye view (BEV) projection of the dose & 56 \\
Bev2Fluence & Dose BEV-to-Fluence network & 18 \\
DL Dose Computation & Predicts dose from fluence maps & 654 \\
Dose Error Construction (\errModule) & Computes 3D dose error map & 1 \\
BEV Projection (2nd pass) & Projects 3D dose error to BEV & 55 \\
Fluence Correction & Network to refines fluence maps & 18 \\
FM CPU to GPU & Transfer the fluence maps from GPU to CPU  & 18 \\
Leaf Sequencing & Converts fluence maps to deliverable MLC sequence & 30 \\
\midrule
\textbf{Total} & End-to-end pipeline time & 889 \\
\bottomrule
\end{tabularx}
\end{table}
\fi


\bibliography{example}

\end{document}